\documentclass[fleqn,usenatbib]{mnras}
\pdfoutput=1
\usepackage{newtxtext,newtxmath}

\usepackage[T1]{fontenc}
\usepackage{ae,aecompl}

\usepackage{graphicx}	
\usepackage{amsmath}	
\usepackage{amssymb}    
\usepackage{subcaption}
\usepackage{mwe}
\usepackage{epsfig}
\usepackage{comment}
\usepackage{color}
\usepackage{longtable}
\usepackage{pdflscape}
\usepackage{upgreek}
\usepackage{float}
\usepackage{color}
\usepackage{lscape}
\usepackage{multirow}
\usepackage{afterpage}
\usepackage{ulem}
\usepackage{morefloats}

\newcommand{\kms}{km~s$^{-1}$}

\newcommand{\unitlum}{erg~s$^{-1}~$}
\newcommand{\Msun}{M$_{\odot}$}

\title[Optical follow-up of iPTF16fnl]{Optical follow-up of the tidal disruption event iPTF16fnl: new insights from X-shooter observations.}

\author[F. Onori et al.]{
F. Onori$^{1,3}$,\thanks{E-mail: francesca.onori@inaf.it }
G. Cannizzaro$^{1,2}$,
P. G. Jonker$^{1,2}$,
M. Fraser$^{4}$,
Z. Kostrzewa-Rutkowska$^{1,2}$,
\newauthor
A. Martin-Carrillo$^{4}$,
S.Benetti$^{6}$,
N. Elias-Rosa$^{7,8}$
M. Gromadzki$^{9}$,
J. Harmanen$^{10}$,
\newauthor
S. Mattila$^{10}$,
M. D. Strizinger$^{11}$,
G. Terreran$^{12}$,
T. Wevers$^{5,2}$.
\\
$^{1}$SRON, Netherlands Institute for Space Research, Sorbonnelaan, 2, NL-3584CA Utrecht, the Netherlands\\
$^{2}$Department of Astrophysics/IMAPP, Radboud University, P.O. Box 9010, 6500 GL Nijmegen, the Netherlands\\
$^{3}$Istituto di Astrofisica e Planetologia Spaziali (INAF), via del Fosso del Cavaliere 100, Roma, I-00133, Italy\\
$^{4}$School of Physics, O'Brien Centre for Science North, University College Dublin, Belfield, Dublin 4, Ireland\\
$^{5}$Institute of Astronomy, University of Cambridge, Madingley Road, Cambridge CB3 0HA, UK\\
$^{6}$INAF - Osservatorio Astronomico di Padova, Vicolo dell’Osservatorio 5, 35122 Padova, Italy.\\
$^{7}$ Institute of Space Sciences (ICE, CSIC), Campus UAB, Carrer de Can Magrans s/n, 08193 Barcelona, Spain\\
$^{8}$ Institut d’Estudis Espacials de Catalunya (IEEC), c/Gran Capit\'a 2-4, Edif. Nexus 201, 08034 Barcelona, Spain\\
$^{9}$Warsaw University Astronomical Observatory, Al. Ujazdowskie 4, 00-478 Warszawa, Poland\\
$^{10}$Tuorla Observatory, Department of Physics and Astronomy, FI-20014 University of Turku, Finland\\
$^{11}$Department of Physics and Astronomy, Aarhus University, Ny Munkegade 120, DK-8000 Aarhus C, Denmark\\
$^{12}$Center for Interdisciplinary Exploration and Research in Astrophysics (CIERA) and Department of Physics and Astronomy,\\
Northwestern University, Evanston, IL 60208
}
\date{Accepted 2019 July 23. Received 2019 July 23; in original form 2019 April 2}

\pubyear{2019}

\begin{document}
\label{firstpage}
\pagerange{\pageref{firstpage}--\pageref{lastpage}}
\maketitle

\begin{abstract}
We present the results from Nordic Optical Telescope and X-shooter follow-up campaigns of the tidal disruption event (TDE) iPTF16fnl, covering the first $\sim$100 days after the transient discovery. We followed the source photometrically until the TDE emission was no longer detected above the host galaxy light. The bolometric luminosity evolution of the TDE is consistent with an exponential decay with e-folding constant t$_{0}$=17.6$\pm$0.2 days. The early time spectra of the transient are dominated by broad \ion{He}{II} $\lambda$4686, H$\upbeta$, H$\upalpha$ and \ion{N}{III} $\lambda$4100 emission lines. The latter is known to be produced together with the \ion{N}{III} $\lambda$4640 in the Bowen fluorescence mechanism. Thanks to the medium resolution X-shooter spectra we have been able to separate the Bowen blend contribution from the broad \ion{He}{II} emission line. The detection of the Bowen fluorescence lines in iPTF16fnl place this transient among the N-rich TDE subset. In the late-time X-shooter spectra, narrow emission lines of [\ion{O}{III}] and [\ion{N}{II}] originating from the host galaxy are detected, suggesting that the host galaxy harbours a weak AGN in its core. The properties of all broad emission lines evolve with time. The equivalent widths follow an exponential decay compatible with the bolometric luminosity evolution. The full-width a half maximum of the broad lines decline with time and the line profiles develop a narrow core at later epochs. Overall, the optical emission of iPTF16fnl can be explained by being produced in an optically thick region in which high densities favour the Bowen fluorescence mechanism and where multiple electron scatterings are responsible for the line broadening.  
\end{abstract}




\section{Introduction}

A tidal disruption event (TDE)  takes place when a star passes close enough 
to a super massive black hole (SMBH) to be disrupted by the black hole's tidal forces \citep{hills75,rees88, phinney89, evans89}.
After the disruption, approximately half of the stellar debris remains bound to the BH in highly elliptical orbits, 
while the other half is expelled in unbound orbits (\citealt{strubbe09, lodato11}). As debris from the disrupted star streams back to the BH, a luminous, short-lived, flare is produced. The transient emission usually peaks in the \textit{ UV} or soft \textit {X}-rays and the evolution of the bolometric luminosity with time is expected to 
follow the bound debris fallback rate, with a powerlaw decline $\propto$ t$^{-5/3}$ on the timescale of months to years (e.g., \citealt{evans89, cannizzo90,rees90, lodato09}). 

In the last decades, a significant number of TDEs has been discovered mostly in $X$-ray and Ultraviolet ($UV$) bands 
(see the reviews of \citealt{gezari12} and \citealt{komossa15}). Luminous, high-amplitude $X$-ray flares from quiescent galaxies, 
consistent with the predictions of the tidal disruption scenario, have first been discovered during the 
\textit {ROSAT} $X$-ray all-sky survey (\citealt{bade96, komossa99, komossagreiner99, greiner00}). 
More recently, similar $X$-ray events have been detected with \textit{ Chandra} and \textit {XMM-Newton} based on dedicated 
searches or serendipitous discoveries (\citealt{esquej07, esquej08, saxton12, auchettl17}). Interestingly, \cite{mattila18} reported the infrared discovery of a TDE in the merging galaxy pair Arp299, in which the optical emission is strongly obscured by dust. In this case, a expanding relativistic jet produced by the accretion of stellar debris on the SMBH has been clearly detected and resolved by radio VLBI observations.

Thanks to the development of wide-field optical transient surveys (such as the All-Sky Automated Survey for Supernovae \cite[ASAS-SN,][]{shappee14},
the Palomar Transient Factory \cite[PTF,][]{Law09},  the Panoramic Survey Telescope and Rapid Response System \cite[PanSTARRS,][]{kaiser02} and the Sloan Digital Sky Survey \cite[SDSS,stripe 82][]{york00}, OGLE-IV Transient Search \cite[]{Wyrzykowski14}, Gaia Science Alerts (\citealt{Hodgkin13}, but see \citealt{Kostrzewa-Rutkowska18} for an independent and systematic search for nuclear transients with Gaia),
an increasing number of TDE has been discovered as luminous blue flares in the nuclei of otherwise quiescent galaxies. Such optically selected TDEs are preferentially found in post-starburst galaxies with no (or weak) current star formation (E+A galaxies, \citealt{arcavi14, french16}).

Detailed photometric and spectroscopic 
follow up studies have been performed for several of these (\citealt{vanvelzen11, gezari12, holoien14, arcavi14,  holoien16a, holoien16b, hung17, blagorodnova18}). 
In general, such transients are characterized by optical spectra with broad ($\sim$10$^{4}$ km s$^{-1}$) \ion{He}{II} $\lambda$4686, H$\upalpha$ and H$\upbeta$ emission lines superimposed on a strong blue continuum. However, optical observations of an increasing sample of TDEs have revealed a number of candidates with observational properties that differ from the classical picture. A continuous sequence of spectral types, from He-dominated to H-dominated through intermediate types with both He and H broad emission lines, covering  a variety  of  values  for  the  He-to-H  line ratios has first been identified by \citep{arcavi14}. Afterwards, prominent metal lines in UV spectra and broad \ion{O}{III} and \ion{N}{III} emission lines in the optical, attributed to the Bowen fluorescence mechanism by \cite{blagorodnova18}, have been detected in some TDEs \cite[]{cenko16, brown18, blagorodnova18}. Very recently, \cite{leloudas19} have detected Bowen lines in the optical spectra of the TDE AT2018dyb, showing that these metal lines are quite common in TDEs and identifying a subclass of N-rich tidal disruption events among the TDE population. Interestingly, \ion{Fe}{II} and \ion{O}{III} emission lines have been detected in the optical spectra of the TDEs AT2018fyk and ASASSN$-$15oi by \cite{wevers19b}, suggesting the existence of the subclass of Fe-rich TDEs. Thus, TDEs appear to be an in-homogeneous class of transients as they show 
different properties in \textit {X}-rays and in the optical spectra. Indeed, most of the optically selected TDEs show no (or very weak) \textit {X}-ray emission and the optical properties, such as for instance the time scale of the evolution of the light curve and the peak luminosity, vary considerably from one TDE to another. 



The emission mechanism behind the observed optical light and spectroscopic features as well as the geometry of the emitting region are still unclear.
Different scenarios have been proposed, including outflows \citep{strubbe09, miller15, metzgerstone16}, emission by shocks from intersecting debris streams 
\citep{piran15, Shiokawa15, bonnerot17}, or an optically thick reprocessing envelope at large radii \citep{guillochon14, roth16, roth18}. 
The diversity of the TDEs observed so far, implies that we are still in the taxonomy phase. Well sampled, multi-wavelength follow-up campaigns with  
a dense coverage of the spectral evolution over the whole flare phase are needed to constrain these models.\\

The nuclear transient iPTF16fnl was discovered on  2016 August 26 (Modified Julian Date [MJD] 57\,626) by the intermediate Palomar Transient Factory (iPTF) survey. The host galaxy is Mrk~0950, an E+A galaxy at $z$=0.016328 and with a luminosity distance $D_{\rm L}$=66.6 Mpc (calculated using $H_{\rm 0}$=69.6 km s$^{-1}$ Mpc$^{-1}$, $\Omega_{\rm M}$=0.29, $\Omega_{\rm \Lambda}$=0.71, in the reference frame of the 3K cosmic microwave background). The transient was classified as a TDE through spectroscopic and photometric observation by \cite{gezari16}.
In the discovery paper, \cite{blagorodnova17} presented photometric and optical spectroscopic data and their analysis cover the first $\sim$60 days after the discovery of the transient, while the $UV$ 
spectroscopic evolution over $\sim$100 days was presented by \cite{brown18}. The lightcurve peak was observed on 2016 August 31 (MJD 57\,632.1) at an absolute magnitude of
 M$_{g}$=$-$17.2 mag, with a luminosity evolution consistent with an exponential decay. The peak bolometric luminosity, inferred from \textit{ UV} and optical photometry by \cite{blagorodnova17}, is $L_{\rm p}\sim 10^{43}$erg s$^{-1}$, an order of magnitude fainter than typically observed in other TDEs \cite[]{vanvelzen11, gezari12, arcavi14, chornock14, holoien14, holoien16a, hung17}. Only a marginal soft \textit{ X}-ray detection of the event at $L_{\rm X}$=2.4$^{+1.9}_{-1.1} \times$10$^{39}$ erg s$^{-1}$ was reported by \cite{blagorodnova17}. 

The spectra of iPTF16fnl resemble those 
of other He and H dominated TDEs although the time evolution is faster, which could be explained by for instance a relatively low mass of the black hole responsible for the disruption \cite[]{blagorodnova17}. 
Indeed, this was measured to be $M_{\rm BH} \sim$3$\times$10$^{5}$ M$_{\odot}$, 
using late-time optical spectra of the host bulge--velocity dispersion and the $M_{\rm BH}$-$\sigma_{\star}$ relation \cite[]{wevers17}.\\

In this work we present results from our optical photometric and spectroscopic follow-up of iPTF16fnl, which started soon after the transient discovery 
as part of the NOT Unbiased Transient Survey (NUTS\footnote{http://csp2.lco.cl/not/}). The monitoring campaign covers $\sim$100 days of the TDE emission and include
medium resolution spectra, obtained with X-shooter under the program 297.B-5062. Thanks to this unique dataset, we have been able to obtain good quality photometric measurements and to perform an accurate spectroscopic analysis, in particular on the broad features observed in the high-resolution X-shooter optical spectra.  
We adopt the date of peak as determined in the optical light curve in the $g^{\prime}$ band by \cite{blagorodnova17}, MJD 57\,632.1, as the reference epoch in all of the following.
We take the foreground (Milky Way) extinction towards Mrk~950 to be $A_V$=0.226 mag \cite[][via the NASA/IPAC Extragalactic Database (NED)]{Schlafly11}.


\section{Observations and data reduction}
\label{sec:obs}
We have monitored the optical emission of iPTF16fnl with both photometric as well as spectroscopic observations, starting from 2016 August 31 and running until 2017 December 16. We used the Andalucia Faint Object Spectrograph and Camera (ALFOSC), mounted on the Nordic Optical Telescope (NOT) on La Palma, Spain and the Asiago Faint Object Spectrograph and Camera (AFOSC), mounted on the Copernico 1.82m telescope in Asiago, Italy. 
In order to have higher resolution and a wide spectral coverage, we have obtained seven spectra over the period from 2016 September 13 to 2016 November 25, using the X-shooter spectrograph \citep{Vernet11}, mounted at the Cassegrain focus of the UT2 at the Very Large Telescope (VLT).
We monitored the photometric evolution with the Watcher telescope \cite[][]{watcherpaper} using Johnson \textit{V} and Sloan $g^\prime$, $r^\prime$ and $i^\prime$ filters. Since the Watcher observations were always very close in time to the NOT observations and the latter are of a higher quality, we decided against using the former.
iPTF16fnl was also followed by the Neil Gehrels \textit {Swift} observatory over a period spanning ~300 days, starting from the transient discovery.
All the spectroscopic observations have been done with the slit oriented at the parallactic angle \cite[]{filippenko82}. In the sections below we describe the spectroscopic and photometric observational set-up and the data reduction for each dataset.  
Throughout this manuscript we report times with respect to 2016 August 31 (MJD 57\,632.1), unless otherwise mentioned.

\subsection {NOT/ALFOSC and Copernico/AFOSC observations}
We obtained iPTF16fnl spectra and images using the ALFOSC instrument over a period of 107 days after the transient discovery. Host--galaxy template images have been obtained on 2017 January 18. A host galaxy spectrum has been taken on 2017 June 16.  
For our spectroscopic observations we used the grism $\#$4, which covers the 3200-9600 \AA\/ wavelength range and provides a resolution of R=$\lambda$/$\Delta\lambda$=360, for a 1.0\arcsec slit under seeing conditions of 1\arcsec~or larger. 

In the framework of the NUTS monitoring campaign, we have obtained two spectra using Copernico/AFOSC instrument on 2016 Sept 09 and 2016 Dec 06, respectively. For the first observation we used grism VPH6, which covers the 4500-10000 wavelength range and provide a seeing limited resolution of R=500, for a 1.0\arcsec slit. The second spectrum has been taken using grism VPH7, which covers the 3200-7000 wavelength range and provide a seeing limited resolution of R=470.  

The details of the NOT/ALFOSC and Copernico/AFOSC spectroscopic observations, such as the observations date, exposure time, the slit width, the airmass, and seeing are shown in Table \ref{tbl:obsSpec}.  

The spectra have been reduced using modfied version of the \texttt{foscgui 1.4} pipeline\footnote{\texttt{foscgui} is a graphical user interface aimed at extracting spectroscopy and photometry obtained with FOSC-like instruments. It was developed by E. Cappellaro. A package description can be found at http://sngroup.oapd.inaf.it/foscgui.html.}, which is based on standard \texttt{iraf} reduction tasks \citep{tody86} and includes bias and flat field correction, cosmic ray cleaning, wavelength calibration using arcs, extraction of the spectra from science frames and flux calibration using a standard star. We measured a FWHM $\sim$ 14 \AA\ for the sky line [\ion{O}{I}] $\lambda$5577 \AA\ in all ALFOSC spectra. Given that in nearly all observations the seeing was of the same order or larger than the slit width, this imply a resolution R$\approx$400. In Figure \ref{fig:NOTspec} we show the sequence of ALFOSC and AFOSC spectra.

Over the same time-period, we also took images of the transient with ALFOSC,  using $u^\prime$, $B$, $V$, $g^\prime$, $r^\prime$, $i^\prime$, $z^\prime$ filters. The image reduction has been performed using the \texttt{foscgui} 1.4 pipeline in imaging mode, which is also based on standard \texttt{iraf} reduction tasks, and include cosmic rays removal, bias and flat fields correction. It also provides the World Coordinate System calibration using SDSS stars. The pipeline gives measurements of the seeing (full-width at half-maximum [FWHM]) and the photometric zero point using SDSS stars in the field of view and the AAVSO Photometric All-Sky Survey (APASS) stars for the Johnson filters.

\begin{figure}
\includegraphics[width=1\columnwidth]{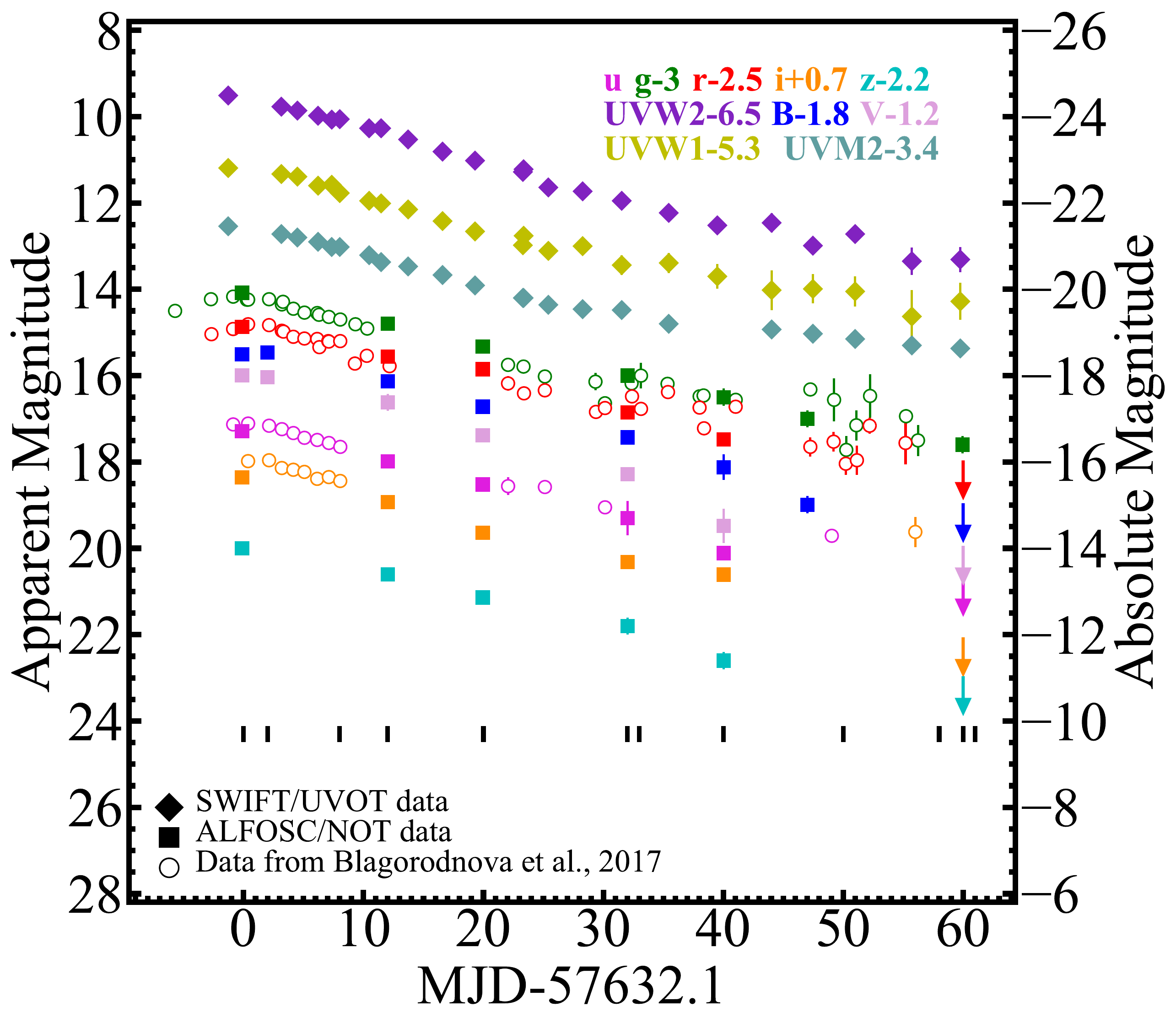}
\caption{Lightcurve for iPTF16fnl. Open circles indicate the optical photometry data from \citet{blagorodnova17}, filled squares indicate the NOT/ALFOSC and UVOT/\textit {Swift} data from this work, arrows indicate the 3$\sigma$ upper limits. The apparent magnitudes are not corrected for foreground Galactic extinction. Magnitudes in Sloan and UVOT/\textit {Swift} filters are in AB system, while magnitudes in Johnson filters are in Vega system. All magnitudes are host-subtracted.
Time zero on the X-axis corresponds to light curve maximum as defined in \citet{blagorodnova17}. Vertical black dashes indicate the times of the spectroscopic observations. 
}
\label{fig:NOTlc}
\end{figure}

For each observation, differential photometry against the host galaxy image has been performed. We have used \texttt{hotpants} V5.1.10 to subtract the host galaxy contribution from the transient images. The software uses the \cite{alard98} algorithm to determine and apply a spatially varying convolution kernel that matches the PSFs of the two images (the transient and the host images) prior to subtraction. In this process, all constant luminosity sources are subtracted and only the (nuclear) transient remains in the subtracted image. Finally, aperture photometry on the subtracted image has been applied using the \texttt{iraf} task \texttt{phot}.
The magnitude uncertainties are calculated by adding in quadrature the standard deviation due to the scatter of the zero point sources and the photometric error on the aperture photometry. No band-pass corrections were applied to the photometric measurements as the uncertainties are dominated by systematic errors due to the template subtraction process. In Table \ref{tbl:phot} we report the measured apparent magnitudes for each NOT/ALFOSC filter, not corrected for foreground extinction and in their common systems (AB for Sloan filters and Vega for Johnson filters). In Figure \ref{fig:NOTlc} the multi-filter NOT/ALFOSC (filled squared in different colors) along with optical photometric data reported in \citet{blagorodnova17} (open circles) are shown. 

\begin{table*}
\centering
 \begin{minipage}{140mm}
 \caption{Spectroscopic observations} 
 \label{tbl:obsSpec}
 \begin{center}
 \begin{tabular}{@{}lclccll}
 \hline
\multicolumn{7}{c}{NOT/ALFOSC observations: Grism \#4; R = 400.}\\
\hline
MJD      & UT Date & phase   & exposure time   & slit   &airmass & seeing \\
         &         & [days]  & [s]             &[\arcsec]  &[\arcsec]  & [\arcsec] \\
(1)      & (2)     &  (3)    & (4)             &(5)     & (6)    &(7)     \\
\hline
57\,631.97 & 2016 Aug\phantom{t} 31 & 0 &  1800         &1.0 &1.35   &  1.19\\ 
57\,634.10 & 2016 Sept 03           & 2 &\phantom{1}900 &1.0 &1.00   &  1.05\\
57\,640.05$^{a}$ & 2016 Sept 09           & 8 & 2700 &1.7 &1.03   &  1.2\\
57\,644.14 & 2016 Sept 13           & 12 &  1800         &1.0 &1.01   &  0.95\\
57\,652.04 & 2016 Sept 21           & 20 &  1800         &1.0 &1.05   &  1.01\\
57\,664.14 & 2016 Oct\phantom{t} 03 & 32 &  1800         &1.0 &1.16   &  0.95 \\
57\,672.12 & 2016 Oct\phantom{t} 11 & 40 &  1800         &1.0 &1.28   &  0.76 \\
57\,692.05 & 2016 Oct\phantom{t} 31 & 60 &  2400         &1.0 &1.21   &  0.90 \\
57\,720.91 & 2016 Nov\phantom{t} 28 & 89 &  1800         &1.0 &1.01   &  0.87 \\
57\,728.90$^{b}$ & 2016 Dec\phantom{t} 06 & 97 & 2400          &1.7 &1.22   &1.50\\
57\,736.90 & 2016 Dec\phantom{t} 14 & 105 &  2400         &1.0 &1.07   &  0.80 \\
57\,738.90 & 2016 Dec\phantom{t} 16 & 107 &  2400         &1.3 &1.07   &  3.62 \\
57\,921.18 & 2017 June 17           & 289 &  2400         &1.0 &1.40   &  1.07 \\
\hline
\multicolumn{7}{c}{VLT/X-shooter observations: R (UVB/VIS/NIR)=6190/11150/8000}\\
\hline
57\,644.25    & 2016 Sept 13           &  12  &590/590/300 &0.8/0.7/0.6 & 1.87& 0.55\\
57\,665.13    & 2016 Oct\phantom{t} 04 &  33  & 590/590/300 &0.8/0.7/0.6 & 1.98& 0.75\\
57\,682.17    & 2016 Oct\phantom{t} 21 &  50  &590/590/300 &0.8/0.7/0.6 & 1.94& 0.80\\
57\,690.05    & 2016 Oct\phantom{t} 29 &  58  &590/590/300 &0.8/0.7/0.6 & 2.09& 0.38\\
57\,692.12    & 2016 Oct\phantom{t} 31 &  60  &590/590/300 &0.8/0.7/0.6 & 1.87& 0.61\\
57\,693.07    & 2016 Nov\phantom{t} 01 &  61  &590/590/300 &0.8/0.7/0.6 & 1.91& 0.64\\
57\,717.04    & 2016 Nov\phantom{t} 25 &  85  &714/714/240 &0.8/0.7/0.6 & 1.86& 0.49 \\
\hline
\end{tabular}
\end{center}
Notes: (1) Modified Julian dates of observations; (2) UT date, (3) Phase with respect to the estimated date of light curve peak MJD 57\,632.1 according to \cite{blagorodnova17}  (4) Exposure time, (5) Slit width  (6) Airmass, (7) Seeing. The reported instrumental resolution (R=$\uplambda$/$\upDelta\uplambda$) has been derived from the measured FWHM of skylines in the spectra.
The X-shooter exposure times and slit widths are reported for UVB, VIS and NIR arms, respectively.\\
(a) Copernico/AFOSC observation,  Grism VPH6, R = 500.\\
(b) Copernico/AFOSC observation,  Grism VPH7, R = 470.
\noindent
\end{minipage}
\end{table*} 

\subsection {Swift satellite UV data}
 The Ultraviolet and Optical Telescope (UVOT) \textit {UVW2}, \textit {UVM2} and \textit {UVW1} images have been reduced using the standard pipeline with the updated calibrations from the \texttt{HEAsoft}-6.24 \texttt{ftools} package, while we used the \texttt{HEAsoft} routine \texttt{uvotsource} to derive the apparent magnitude of the transient. The aperture photometry has been measured using a 5\arcsec\/ aperture centered on the position of the transient and  a background region of 60\arcsec\/ radius placed in an area free of sources. In Table \ref{tbl:UVphot} we provide the measured UVOT/\textit {Swift} \textit {UVW2}, \textit {UVM2} and \textit {UVW1} apparent magnitudes in the AB system \cite[zero points from][]{poole08} and the corresponding flux densities. The magnitude values reported in Table \ref{tbl:UVphot} are not corrected for foreground extinction. 
 Swift/UVOT suffers from a well-known ``red leak'' of optical photons to the UV bands. However, we do not expect this to significantly affect our results, based on \cite{brown16}, who found at most a 20\% or 5\% effect on photometry in the \textit {UVW2} and \textit {UVW1} bands respectively.
 
Even in the late-time images (MJD 57\,932.05 $\sim$300 days from the transient discovery) there is still emission. We ascribe this detection to host--galaxy light as there is no TDE emission detection in our latest NOT/ALFOSC optical images, which were taken at MJD 57\,771. Moreover, spectroscopic signatures of the TDE are absent in the NOT/ALFOSC spectrum taken at MJD 57\,921.18.
In order to derive the TDE \textit {UVW2}, \textit {UVM2} and \textit {UVW1} magnitudes, we subtracted the host contribution as determined from the late time images (MJD 57\,932.05) from the photometric measurements. 
In Figure \ref{fig:NOTlc} the host subtracted UVOT/ \textit {Swift} \textit {UVW2}, \textit {UVM2} and \textit {UVW1} apparent magnitudes (filled diamonds in different colors), together with the ALFOSC data are shown.

\subsection{VLT/X-shooter spectra}
We have obtained a total of 7 spectra using the VLT/X-shooter spectrograph. X-shooter spectra cover the spectral range from 3000 to 25000 \AA. The observations span the period from 2016  September 13 to 2016 November 25 and have been carried out using slit widths of 0\farcs8, 0\farcs7 and 0\farcs6 for the UVB, VIS and NIR arms, respectively. The length of the slit is always 11\arcsec. This set-up yields a seeing limited resolution of R = 6190, 10640, and 8040 for the UVB, VIS, and NIR arm, respectively. All X-shooter spectra have been taken with the slit oriented along the parallactic angle. The details of the observations, including the instrumental configuration, the exposure times, and the observing conditions are reported in Table~\ref{tbl:obsSpec}. The data reduction has been performed using the \texttt{reflex} X-shooter pipeline 2.8.0 \cite[]{freudling13}, while the correction for atmospheric absorption features has been done using the software \texttt{molecfit}, which fits synthetic transmission spectra to the astronomical data \cite[]{smette15,kausch15}. We measured a FWHM$\sim$0.5 \AA\ for the unsaturated sky line [\ion{O}{I}] $\lambda$5577 in all the VLT/X-shooter VIS spectra and a FWHM$\sim$2 \AA\ for sky lines around $\lambda$16000 \AA\ in all the NIR spectra, in order to derive the instrumental resolution in these wavelength regions. In Figure \ref{fig:NOTspec} the VLT/X-shooter UVB and VIS spectra sequence is shown, while the VLT/X-shooter NIR spectra are shown in Figure \ref{fig:NIRspec}. No significant spectroscopic features have been found in the NIR part of the spectrum.

\section{Photometric evolution}
\label{sec:photometry}
From the UVOT/\textit {Swift} and NOT/ALFOSC host subtracted apparent magnitudes ($UVW2$, $UVM2$, $UVW1$, $u^\prime$, $B$, $V$, $g^\prime$, $r^\prime$, $i^\prime$, $z^\prime$) we compute the pseudo bolometric luminosity for iPTF16fnl using the \texttt{python} routine \texttt{superbol} \cite[]{nicholl18}.
In order to cover the optical pre-peak phase, we also include the $g^\prime$ and $r^\prime$ measurements from \citet{blagorodnova17}.
All the input magnitudes have been corrected for the Galactic extinction from \cite{Schlafly11} which assume a reddening law with Rv = 3.1 and K-correction has been applied. 

\begin{figure}
\includegraphics[width=1\columnwidth]{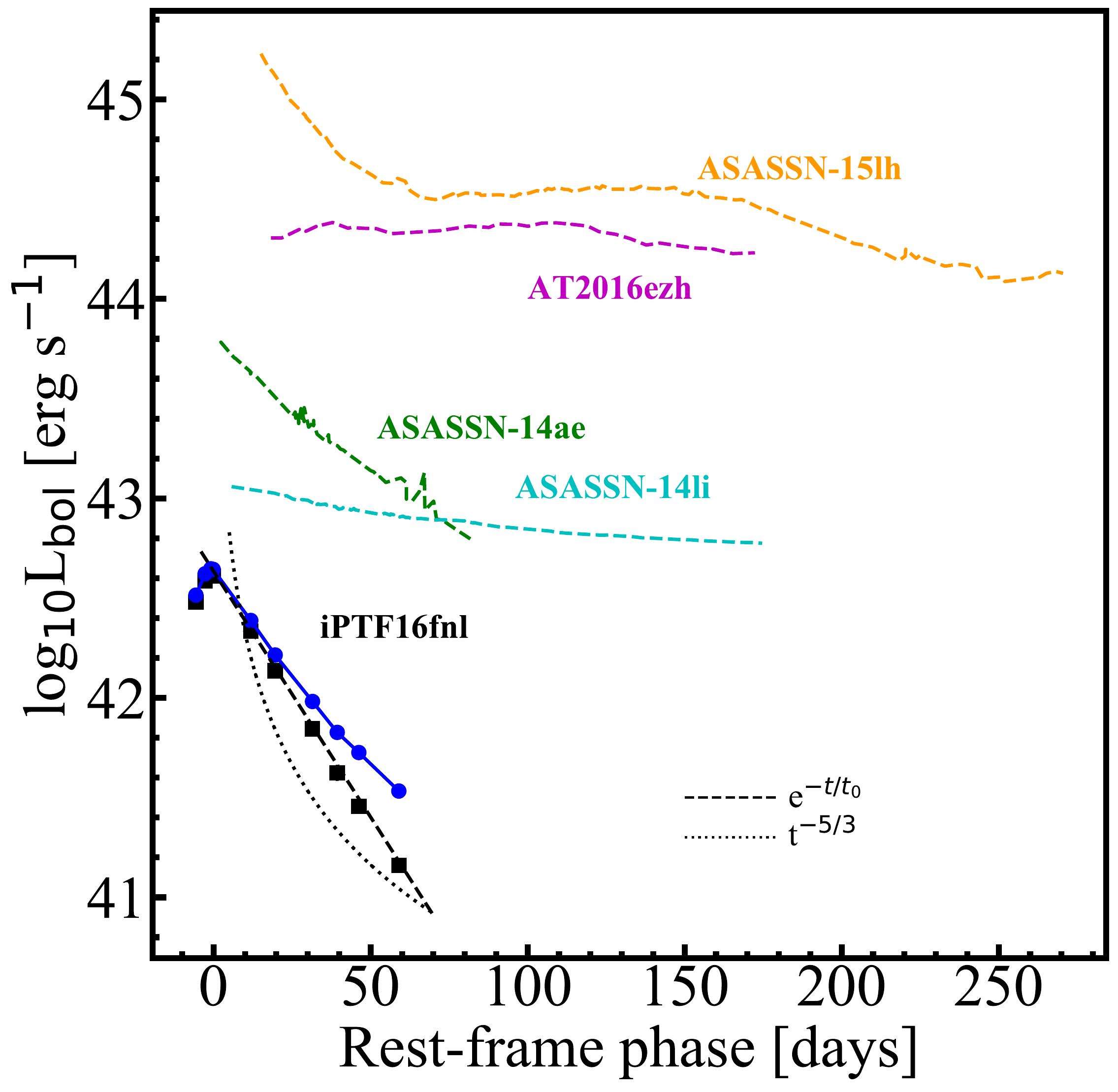}
\caption{iPTF16fnl bolometric light curve obtained from integration of observed fluxes (black filled squares). The blue filled circles, connected by a solid blue line show the bolometric light curve obtained using no host subtracted UV fluxes. For comparison we report the bolometric light curves for some TDEs: ASASSN-14li (\citet{holoien16a}, cyan dashed line), ASASSN-14ae (\citet{holoien14}; green dashed line), AT2016ezh (\citet{blanchard17}; magenta dashed line) and ASASSN-15lh (\citet{Leloudas16}, but see also \citet{dong16} and \citet{godoy17} for the SLSN-I interpretations for this transient; yellow dashed lines). Black dashed and dotted line indicate the exponential and the t$^{-5/3}$ decline for iPTF16fnl, respectively. Our best fit is obtained with L$\propto e^{t/t_{0}}$ with t$_{0}$=17.6$\pm$0.2.}
\label{fig:lb}
\end{figure}

In Figure \ref{fig:lb} the iPTF16fnl bolometric light curve, obtained by integrating flux over observed filters is shown. For comparison, we show the pseudo bolometric luminosity evolution for some TDEs, obtained using UV and optical bands: ASASSN-14li (\citet{holoien16a}, cyan dashed line), ASASSN-14ae (\citet{holoien14}; green dashed line), AT2016ezh (\citet{blanchard17}; magenta dashed line) and ASASSN-15lh (\citet{Leloudas16}, but see \cite{dong16} and \cite{godoy17} for the SLSN-I interpretation on this transient; yellow dashed lines).
The luminosity evolution of iPTF16fnl is remarkably fast with a faint bolometric luminosity, which at peak is $L_{\rm p}$ $\sim$ (4$\pm$1)$\times$10$^{42}$ erg s$^{-1}$. The luminosity decline is well fitted by the exponential model $L_{\rm bol} \propto$e$^{-t/t_{0}}$, with an e-folding constant $t_{\rm 0}$=17.6$\pm$0.2, while the usual powerlaw function cannot represent the data (see Figure \ref{fig:lb} for a comparison between the two models).
The total energy radiated ($E_{\rm rad}$) is derived by integrating the bolometric luminosity over time and it results $E_{\rm rad}$= (8.07$\pm$0.78)$\times 10^{48}$ erg. 

Recently, \cite{vanvelzen18} reported the detection of long-lived $UV$ emission ascribed to the TDE accretion disk. In order to estimate the contamination to the iPTF16fnl light curve if the late-time $UV$ measurements are due to the TDE accretion disk and not to the host galaxy, we derived the bolometric luminosity evolution also using the $UV$ filters fluxes without applying the host subtraction procedure. In Figure \ref{fig:lb} the comparison between the bolometric luminosity derived with the two methods is shown. At early times the the two bolometric luminosities are consistent with being the same. Instead, the difference between the bolometric luminosities as determined using the two methods becomes larger at late times. In the last epoch the value for the $L_{\rm bol}$ obtained using the $UV$ fluxes without the host subtraction is a factor of three higher than the value obtained in the case of host subtracted $UV$ fluxes. It also results in a higher total energy radiated, which is $E_{\rm rad}$= (9.56$\pm$0.68)$\times 10^{48}$ erg.

\begin{figure}
\includegraphics[width=1\columnwidth]{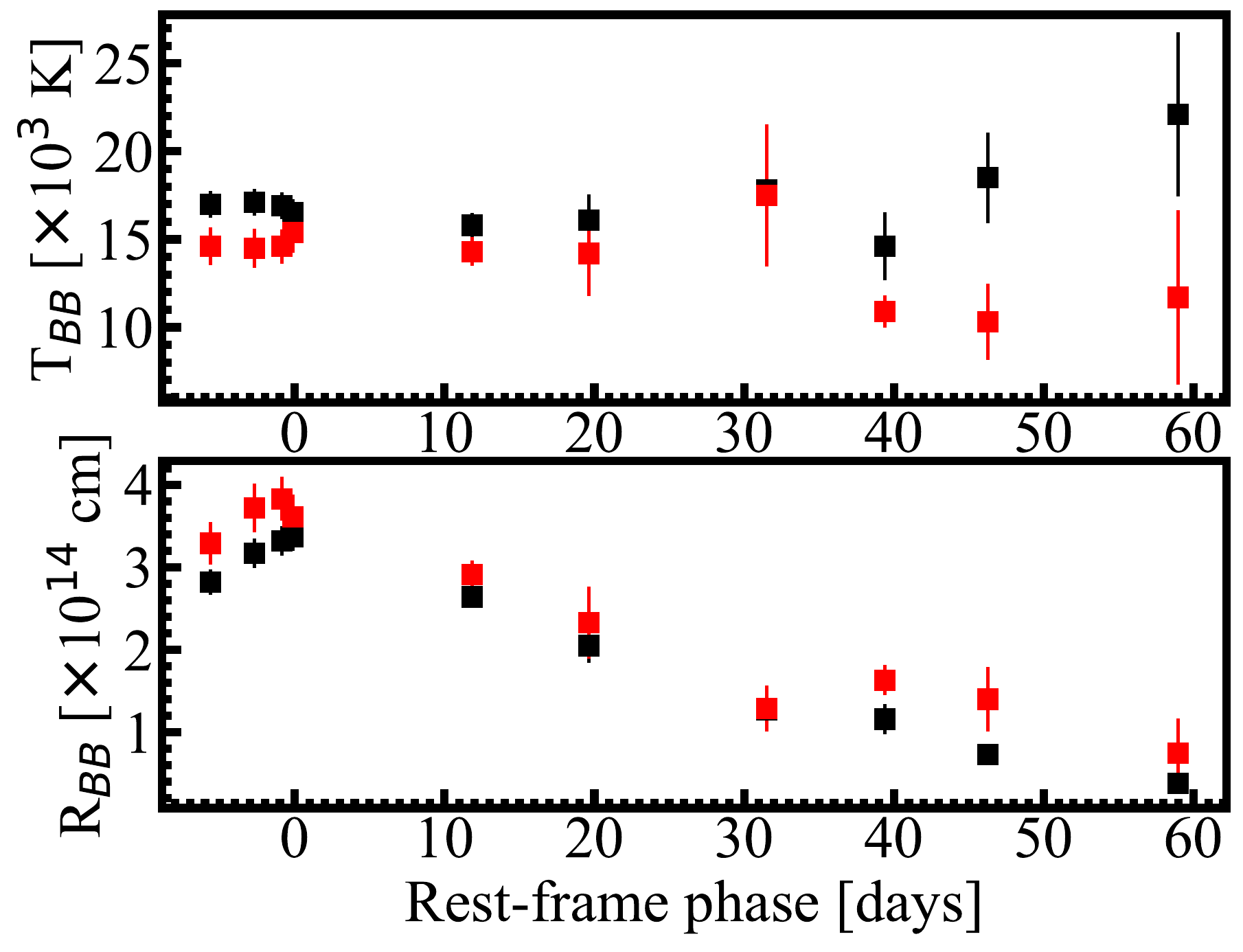}
\caption{Evolution of temperature ($T_{\rm BB}$) and radius ($R_{\rm BB}$) obtained from black body fits to the observed spectral energy distributions (SED). Black filled squares: values obtained fitting a single black body to the SED using the fluxes at all wavelengths. Red filled squares: values obtained by fitting a black body to the optical SED data only.}
\label{fig:bb}
\end{figure}

\begin{figure}
\includegraphics[width=1\columnwidth]{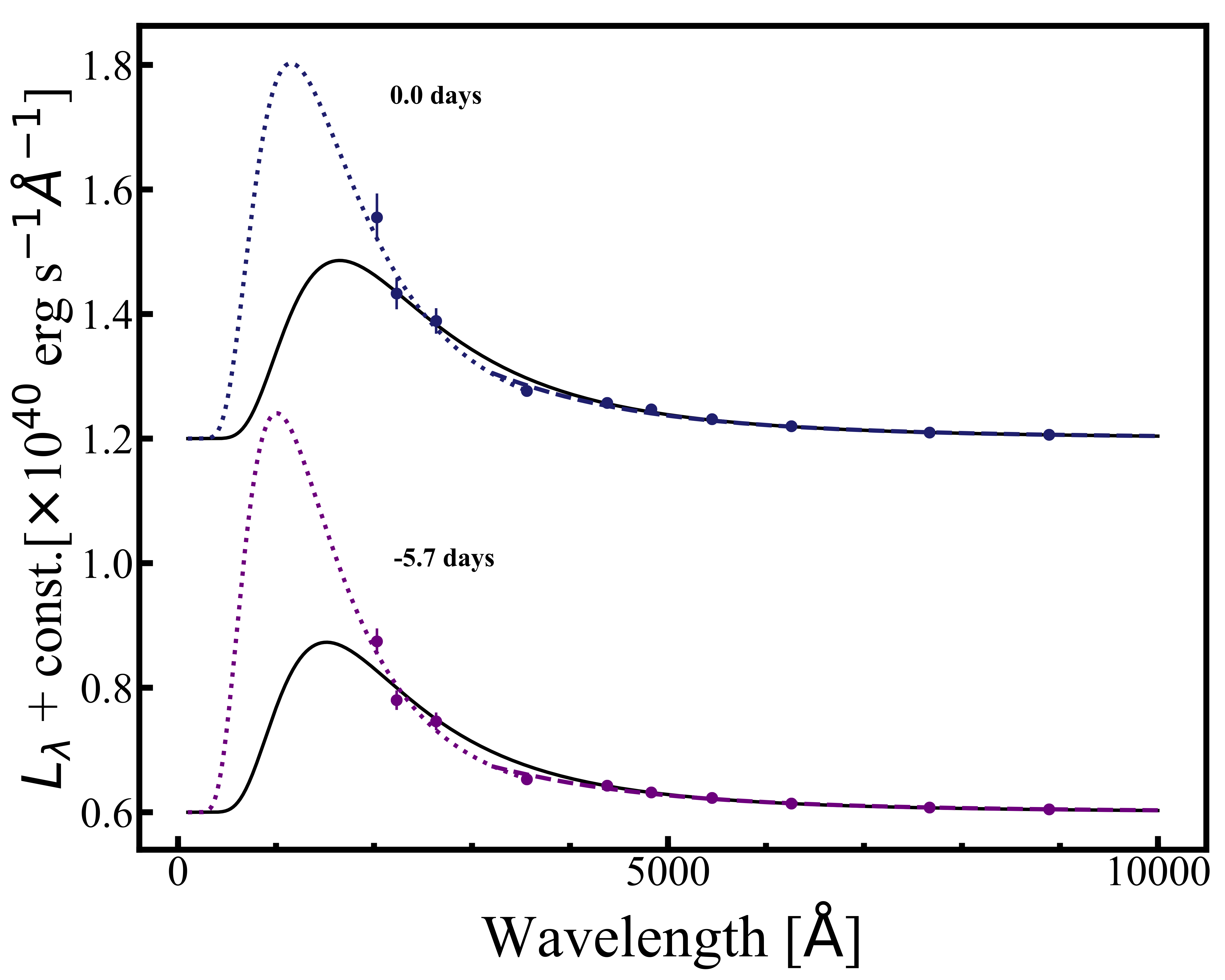}
\caption{Fit of the black body models to the iPTF16fnl luminosity density for the two epochs of observations (phase=-5.7 days and phase=0, blue and purple, respectively). The colored points correspond to the luminosity values for each filter, while the lines are the BB models. In particular, the black solid line is the BB model applied to all wavelengths, while the colored dotted and dashed lines are the BB model applied only to UV filters and the BB model applied only to optical data, respectively.}
\label{fig:bbfits}
\end{figure}

In Figure \ref{fig:bb} the evolution of the black body (BB) radius, $R_{\rm BB}$ and BB temperature, $T_{\rm BB}$, obtained from two different fitting methods are shown. In particular, the black--filled squares show the evolution of $R_{\rm BB}$ and $T_{\rm BB}$ obtained by fitting all wavelengths with a single black body. The red--filled squares show the results from the black body fit to only the optical data. As an example, we show in Figure \ref{fig:bbfits} the fit of the BB models to the luminosity density for two epochs of observations. 

While the evolution of $R_{\rm BB}$ does not change significantly between the two methods, showing a similar decline with time, when we take into account also the UV luminosity, the black body temperature $T_{\rm BB}$ increases from $\sim$40 days after the peak and it reaches the maximum value of (22.1$\pm$4.7)$\times$10$^{3}$ K. However, we note that the optical fit gives more reliable results for the temperature, as it exclude the line-blanketed region in UV band \cite[see][]{nicholl17, yan18}. In particular, the BB temperature obtained by using the optical BB fit is consistent with being at a constant value of $T_{\rm BB} \sim$1.5$\times$10$^{4}$ K at all phases. Instead, the evolution of $R_{\rm BB}$ is characterized by an initial expansion during the epochs prior to the TDE peak, when it reaches its maximum value of $R_{\rm BB}$=(3.6$\pm$0.3)$\times$10$^{14}$ cm, followed by a declining trend. After $\sim$60 days it reaches a minimum value of $R_{\rm BB}$=(0.8$\pm$0.4)$\times$10$^{14}$ cm. We note that this value is $\sim$three orders of magnitude larger than what is expected for a Schwartzchild radius derived for a black hole mass of $M_{\rm BH}\sim$ 3$\times$10$^{5}$\Msun, as measured by \cite{wevers17}, and $\sim$one order of magnitude larger than the expected tidal radius for the disruption of a solar-like star. This suggests that the optical emission of iPTF16fnl could be produced in a region at a distance larger than the tidal radius.  
Interestingly, the origin of the optical emission for a sample of optically selected TDEs has been investigated by \cite{wevers17}. In their work, the authors found similar results for the black body radii of the TDEs in their sample and conclude that stream self-intersection or accretion--powered re--processing models can explain the observed \textit{UV}/optical TDE emission.  

\begin{figure}
\includegraphics[width=1\columnwidth]{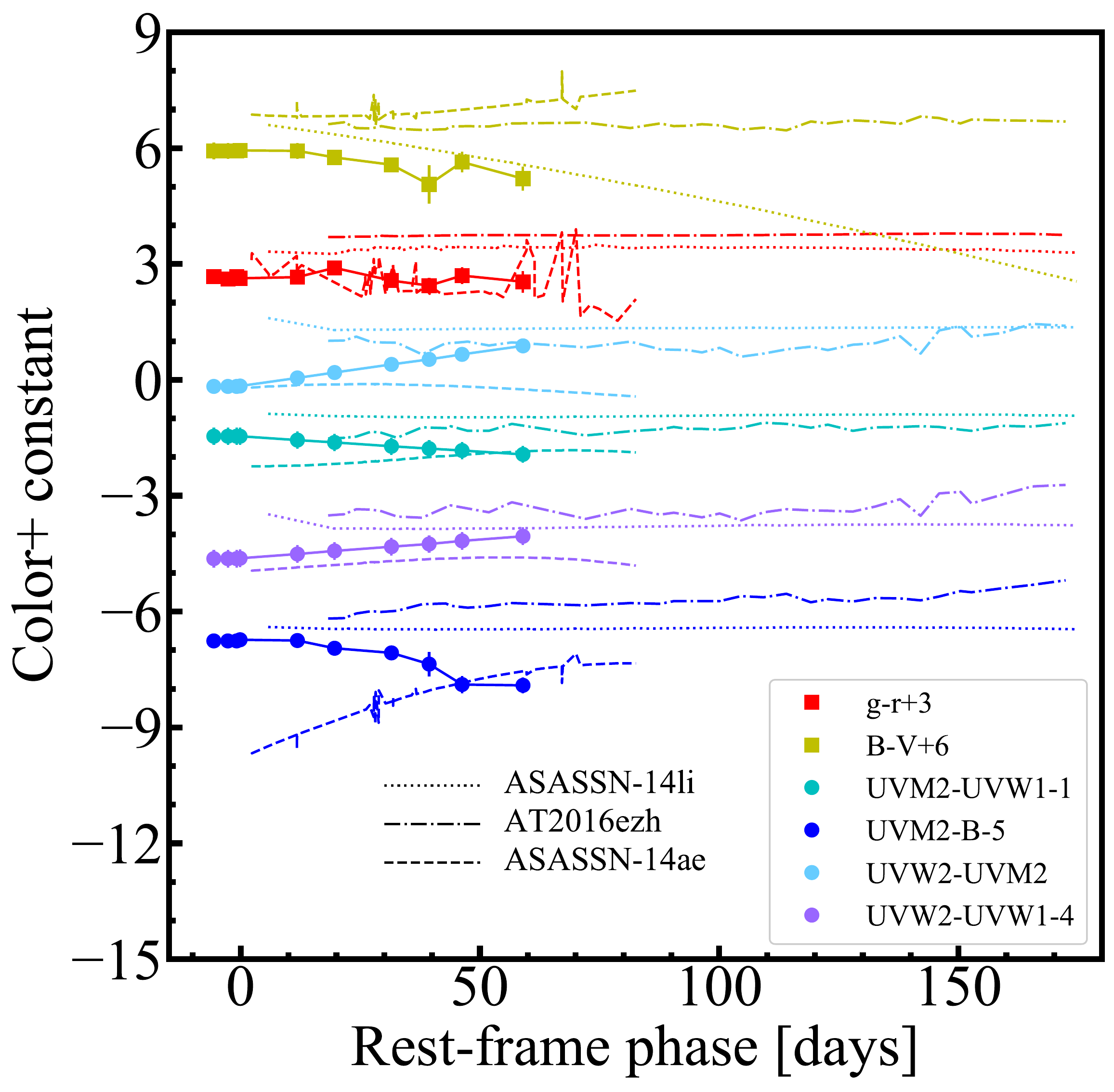}
\caption{UV and optical color evolution for iPTF16fnl (filled squares for optical colors and filled circles for UV colors) compared with three TDEs: ASASSN-14li (\citet{holoien16a}, dotted lines), ASASSN-14ae (\citet{holoien14}, dashed lines) and AT2016ezh (\citet{blanchard17}; dotted-dashed lines). K-correction has been applied in all data used. The iPTF16fnl data are corrected for Galactic extinction and host-subtracted.}
\label{fig:color}
\end{figure}

In Figure \ref{fig:color} we show the iPTF16fnl UV and optical color evolution for the following filters: \textit {UVM2}-\textit {UVW1}, \textit {UVM2}-\textit {B}, \textit {UVW2}-\textit {UVM2}, \textit {UVW2}-\textit {UVW1}, \textit {B}-\textit {V} and \textit {g$^{\prime}$}-\textit {r$^{\prime}$}. 

All the data used are host-subtracted and extinction correction, as well as K-correction have been applied. For comparison we also show the colors evolution for the TDEs ASASSN-14li \citep{holoien16a}, ASASSN-14ae \citep{holoien14} and AT2016ezh \citep{blanchard17}. While \textit {g$^{\prime}$}-\textit {r$^{\prime}$} does not show any significant change over the time and shows a behaviour similar to what observed in the the other TDEs reported here, some evolution is visible in all the other colors. In particular, while \textit {B}-\textit {V}, \textit {UVM2}-\textit {UVW1} and \textit {UVM2}-\textit {B} show a decrease with time and become redder, both \textit {UVW2}-\textit {UVW1} and \textit {UVM2}-\textit {UVW1} colors increase with time, becoming bluer. While the \textit {B}-\textit {V} evolution is similar to the behavior observed for ASASSN-14li (yellow dotted line), the other colors trends are quite different with respect to the color evolution observed in the other TDEs reported here for comparison. 

\section{Spectral analysis}
The iPTF16fnl optical spectra show typical features observed in the H+He rich TDEs. In particular, the early spectra (in Figure \ref{fig:NOTspec}) are dominated by a strong blue continuum with broad \ion{He}{II} $\lambda$4686 and H$\upalpha$ emission lines clearly visible. The strong host galaxy contribution is also visible through the sequence of Balmer absorption lines around 4000 \AA, which are typical spectral features of E+A galaxies. In order to accurately identify and analyze the TDE emission lines, it is necessary to subtract the host galaxy contribution from all the spectra. For this purpose, we have taken a late--time spectrum with NOT/ALFOSC on 2017 June 17, 289 days after the discovery, when the TDE emission does not contribute to the optical host light significantly. Indeed, already after 60 days from the light-curve peak, the TDE emission was no longer detected in the optical images (at this phase only upper limits are available, see Table \ref{tbl:phot}). Moreover, no spectroscopic features of the transient emission are present in the last NOT/ALFOSC spectra shown in Figure \ref{fig:NOTspec}.

As spectral signatures of the TDE are still present in our latest X-shooter spectrum, we have used high-resolution stellar templates to obtain a synthetic spectrum of the host galaxy in the X-shooter UVB and VIS arms.  
Before the host subtraction, the reduced spectra have been corrected for the foreground extinction using the Cardelli function \cite[]{cardelli89} with A$_V$=0.226 mag and R(V)=3.1.

\subsection{Host--galaxy subtraction}
In order to take the different observing conditions of each spectrum into account, we have used the penalized pixel fitting (\texttt{ppxf}) method \cite[]{cappellari04,cappellari17} to convolve the host galaxy spectrum to the host+TDE spectra. 

The method approximates the observed galaxy spectrum by convolving a template spectrum $T$(x) (or a series of n templates) by the line of sight velocity dispersion function $f$(v) (LOSVD). The galaxy model is obtained following the general approximation: 

\begin{equation}
\begin{split}
G_{mod} \left(x \right) = {\textstyle \sum_{n=1}^{N}} w_n \{ \left[ T_n \left(x\right) \ast f_n \left(cx \right)\right]{\textstyle \sum_{k=1}^K} a_k \mathcal{P}_k \left(x \right)\} + \\
{\textstyle \sum_{l=1}^L} b_1 \mathcal{P}_l \left( x \right) + {\textstyle \sum_{j=1}^J} c_j S_j \left(x \right),
\end{split}
\end{equation}

\noindent where the $w_n$ are the spectral weights, the $\mathcal{P}_k$ and $\mathcal{P}_l$ are are multiplicative and additive orthogonal polynomials (of Legendre type or a truncated Fourier series), respectively, and $S_j$ are the spectra of the sky. Both polynomials and sky are optional components in the \texttt{ppxf} fit. The LOSVD function (f$\left( cx\right)$= $f\left(v\right)$) is parametrized by a series of Gauss-Hermite polynomials:

\begin{equation}
f\left(v\right) = \frac{1}{\sigma \sqrt{2\pi}} exp \left( \frac{1}{2} \frac{\left( v-V\right)^2}{\sigma^2}\right) \left[ 1 + {\textstyle \sum_{m=3}^M} h_m H_m \left( \frac{v-V}{\sigma}\right) \right],
\end{equation}

\noindent where $V$ is the mean velocity along the line of sight, $\sigma$ is the stellar velocity dispersion, $H_m$ and $h_m$ are the Hermite polynomials and their coefficients, respectively.
The best fitting template is then found by $\chi^{2}$ minimization. 

In the case of NOT/ALFOSC and Copernico/AFOSC dataset, we used as template the NOT/ALFOSC host galaxy spectrum, taken on 2017 June 17 with the same observational set-up as all the previous observations (see Table \ref{tbl:obsSpec}). The host galaxy spectrum is shown in orange in Figure \ref{fig:NOTspec}. There is no sign of TDE emission lines anymore, moreover, the photometry measurements on the latest image we have, taken on the 2017 January 18, indicate that the transient has already faded beyond detection in the optical bands at this time. Indeed, from the apparent optical magnitudes derived from aperture photometry, listed in Table \ref{tbl:phot}, only upper limits are measured in the last epoch.

In the fitting procedure, in order to model both the host galaxy and the TDE blue continuum contributions, we used a 4th degree additive Legendre polynomial. We excluded from the fit all the spectral regions where the broad TDE emission components were present and the regions affected by telluric absorption. 

\begin{figure}
\includegraphics[width=1\columnwidth]{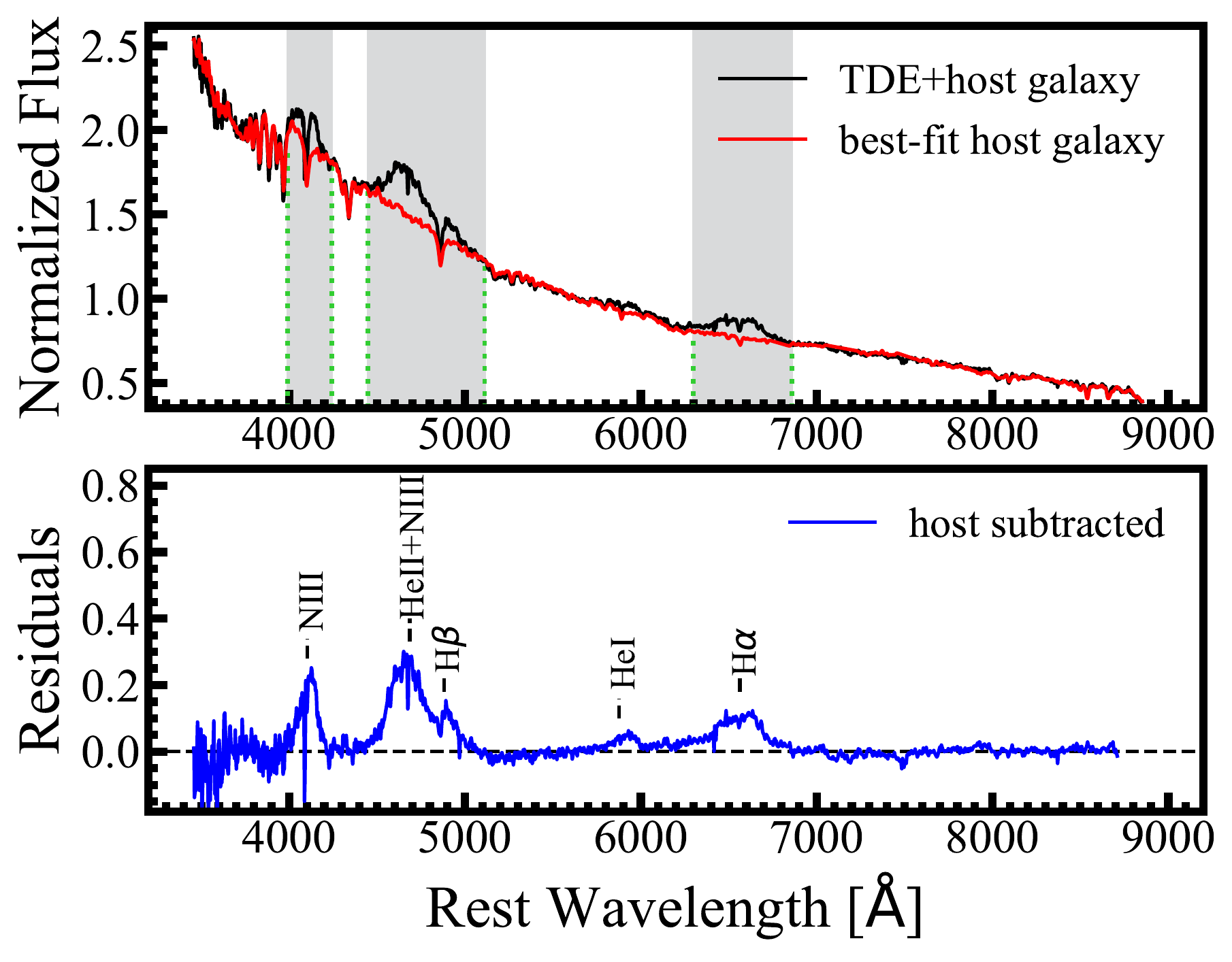}
\caption{Host and TDE continuum subtraction on the NOT/ALFOSC spectrum, taken on 2016 August 31. In the upper panel, the best-fit host galaxy+TDE continuum (in red) is plotted over the TDE+host spectrum (in black). The green dotted lines and the areas in gray indicate the spectral regions excluded during the fit. In the lower panel, the resulting host-subtracted and continuum-subtracted spectrum (in blue) is shown.}
\label{fig:NOThostsub}
\end{figure}

In Figure \ref{fig:NOThostsub}, the subtraction of the host galaxy and the TDE continuum contributions for the NOT/ALFOSC spectrum taken on 2016 Aug 31 is shown. For comparison, the best fit (host galaxy and TDE continuum, in red) is plotted over the normalized host+TDE spectrum (in black). The spectral region excluded from the fit (i.e. the \ion{He}{II} and H$\upalpha$) is indicated with dotted green vertical lines.
The resulting host-subtracted and continuum-subtracted spectrum (in blue) along with the identification of the broad emission lines is also shown. 

In the case of the VLT/X-shooter data-set, we first have produced a synthetic host galaxy template by applying the \texttt{ppxf} method on the latest X-shooter spectra we have (taken the 2016 November 25) using the \texttt{phoenix} v16 high-resolution synthetic spectra \citep{husser13}\footnote{http://phoenix.astro.physik.uni-goettingen.de/}.  

We choose this particular synthetic spectral library for the wide wavelength coverage (from 500 \AA \, to 5.5 $\micron$) and the high resolution (R$\sim$50000 in the range 3000$-$25000 \AA), which are well suited for the X-shooter spectral properties. The whole \texttt{phoenix} library contains $\sim$ 30000 synthetic spectra and covers the properties of most stellar populations: the effective temperature varies between 2300 K $\leq$T$_{eff}\leq$ 12000 K, the metallicity range is $-$4.0$\leq$[Fe/H]$\leq$1.0 and the alpha element abundances ranges between $-$0.2$\leq$[$\alpha$/Fe]$\leq$1.2. We selected a subsample of stellar spectra containing $\sim$5800 spectra with 2300 K $\leq$T$_{eff}\leq$ 12000 K, $-$4.0$\leq$[Fe/H]$\leq$1.0 and [$\alpha$/Fe]=0.

In order to avoid the TDE contamination in the fit of the syntetic host galaxy, the spectral wavelength range of the TDE emission features has been excluded from the fit.
The resulting best-fitting host galaxy synthetic spectrum has been used as single template when applying the \texttt{ppxf} on the remaining X-shooter spectra.\\     

\begin{figure}
\includegraphics[width=1\columnwidth]{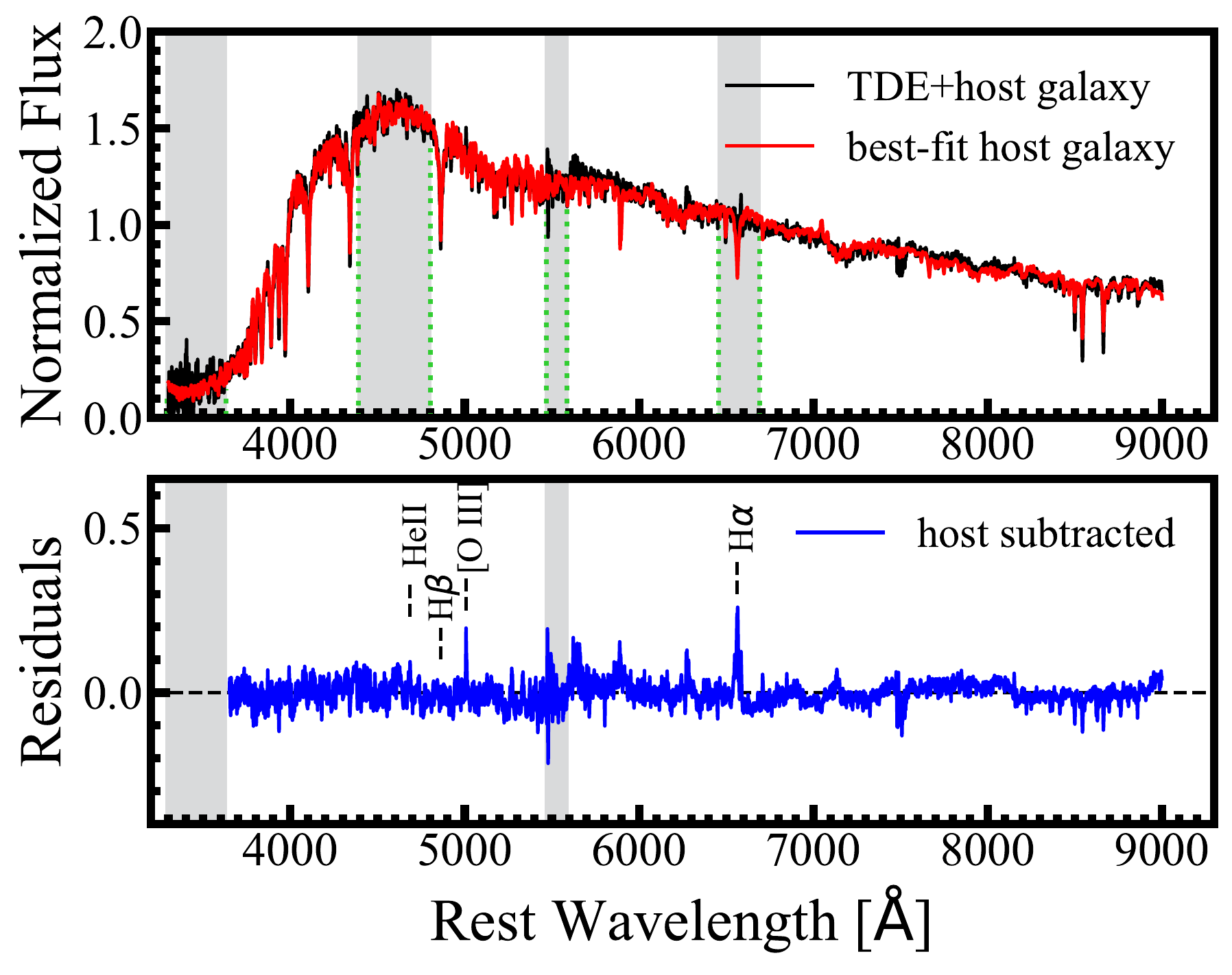}
\caption{Host subtraction on the VLT/X-shooter spectrum, taken on 2016 November 25. In the upper panel, the best-fit host galaxy (in red) is plotted over the TDE+host UVB+VIS spectrum (in black). The green dotted lines and the areas in gray indicate the spectral regions excluded during the fit and the noisy regions due to the overlapping of the UVB and VIS arms, which we also excluded from the fit. In the lower panel, the resulting host-subtracted spectrum (in blue) is shown. We indicate the position of the main emission lines.}
\label{fig:XSHhostsub}
\end{figure}

The host galaxy fit has been applied on both the VLT/X-shooter UVB and VIS spectra in the wavelength region between 3300 \AA \,$-$9000 \AA\ , to avoid the noisy regions at the edges of the spectra. Similarly for the NOT/ALFOSC dataset, we have excluded all the region of the TDE broad emission features from the fit (i.e. the \ion{He}{II} and H$\upalpha$).
In Figure \ref{fig:XSHhostsub} the host galaxy subtraction for the VLT/X-shooter spectrum taken on 2016 November 25 is shown. The best--fit host galaxy spectrum is shown in red and it is plotted over the UVB+VIS TDE+galaxy spectrum, in black. The resulting host-subtracted spectrum (in blue) and the identification of the main emission lines is shown in the lower panel. The vertical gray areas show the spectral regions excluded from the fit.

\section{The TDE emission lines}
\begin{figure}
\includegraphics[width=1\columnwidth]{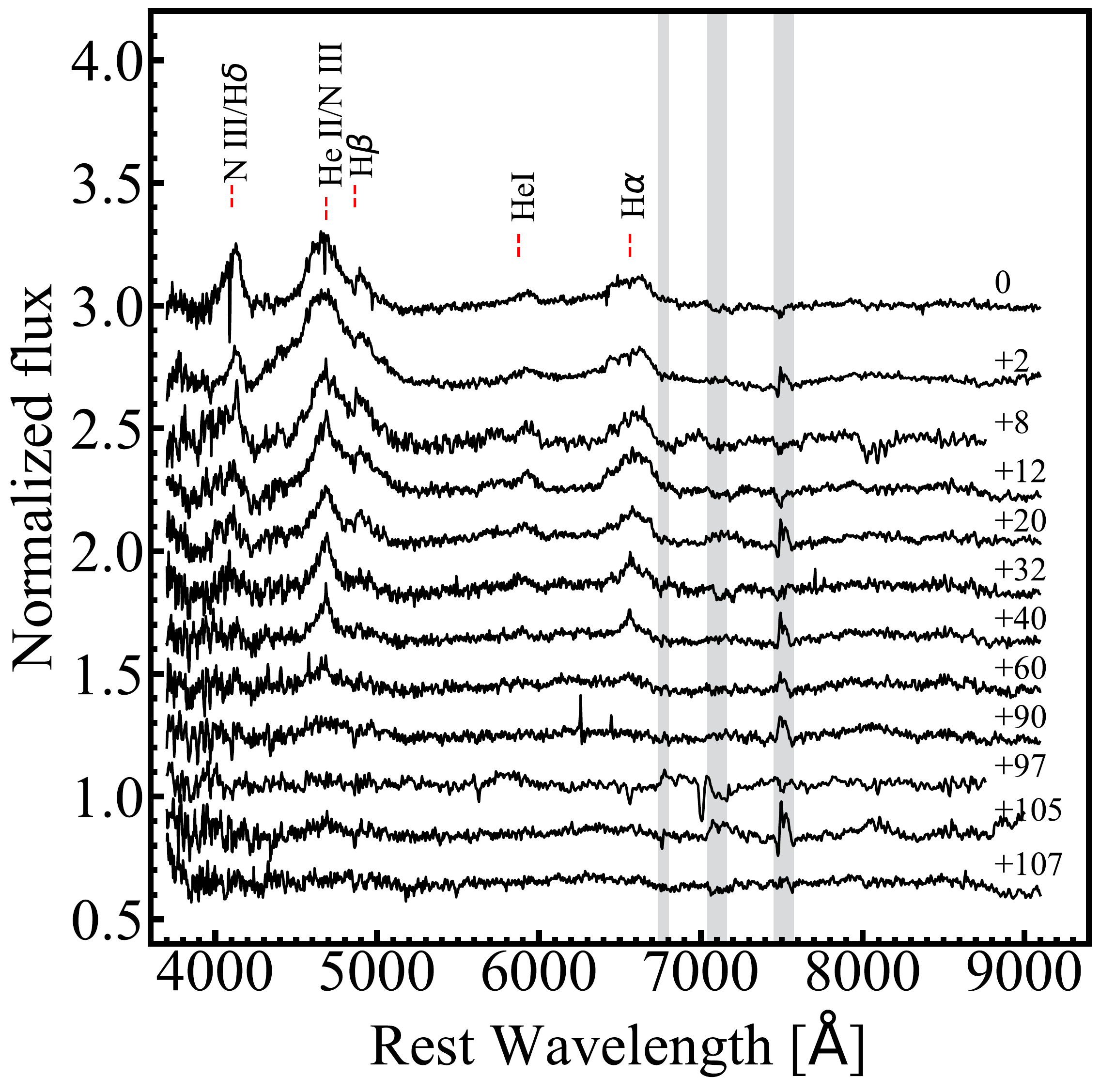}
\caption{Sequence of NOT/ALFOSC and Copernico/AFOSC host-subtracted spectra. The Copernico/AFOSC set of data is made by two spectra taken at phases +8 and +97, respectively.  The main emission lines and their identifications are indicated (vertical dashed red lines). 
Broad components in the \ion{N}{III} $\lambda$4100, \ion{He}{II} $\lambda$4686, H$\upbeta$, \ion{He}{I} and H$\upalpha$ are clearly visible
The gray area indicate the position of telluric lines. All spectra have been normalized to 1, but are shown with offsets for clarity.}
\label{fig:NOTspec_suball}
\end{figure}

\begin{figure}
\includegraphics[width=1\columnwidth]{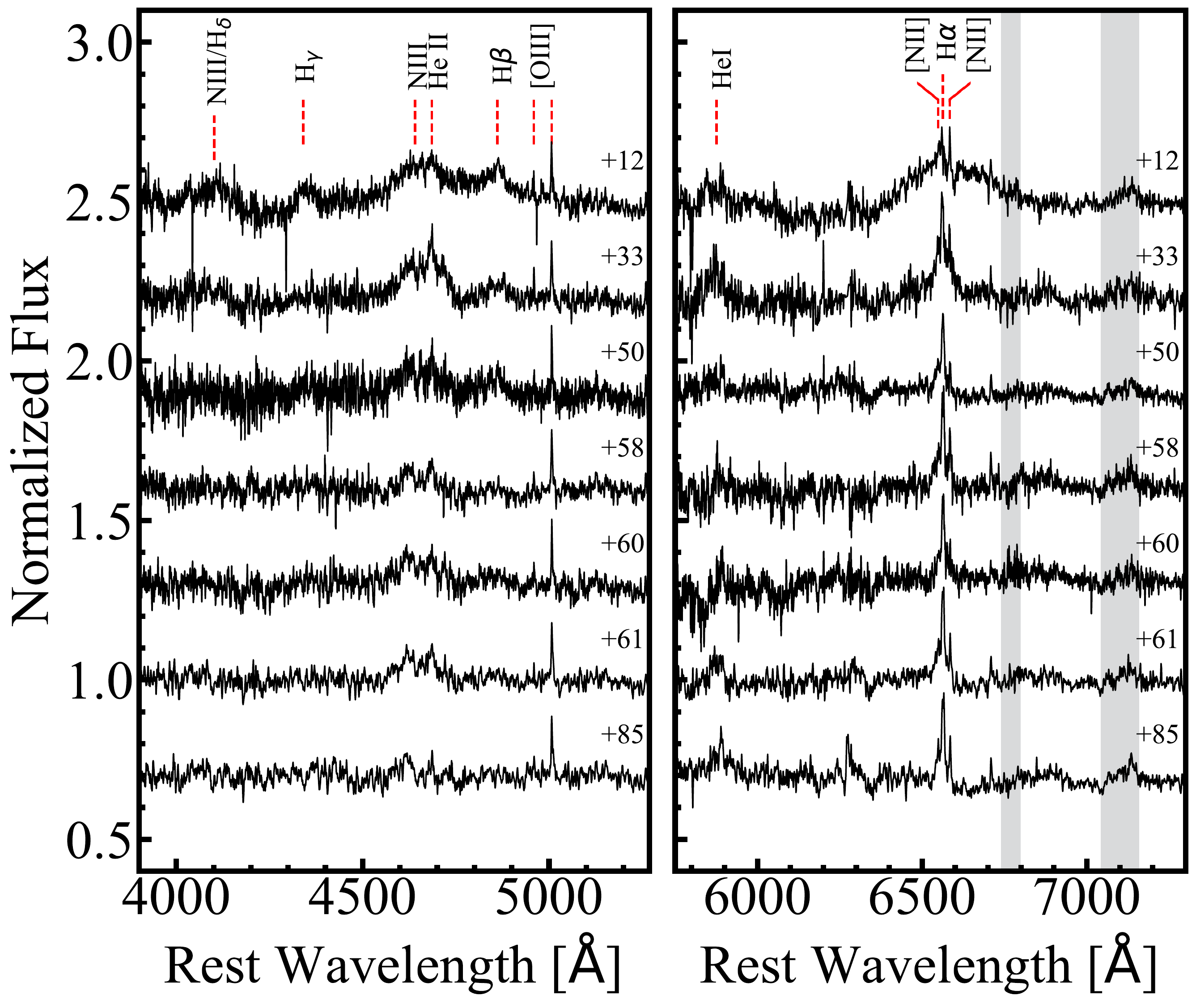}
\caption{Sequence of VLT/X-shooter host-subtracted spectra. In the left panel the \ion{He}{II} region in the UVB spectra is shown, while the H$\upalpha$ area in the VIS spectra is shown in the right panel. The identification of the main emission lines is indicated (vertical dashed red lines). The \ion{N}{III} $\lambda$4640 Bowen blend is evident in the \ion{He}{II} broad feature, which show a double--peaked profile at all epochs. Besides the broad \ion{He}{II} $\lambda$4686 and H$\upalpha$ features, narrow \ion{O}{III} $\lambda$5007 and \ion{N}{II} $\lambda$6583 emission lines are also detected. 
}
\label{fig:XSHspec_suball}
\end{figure}

The sequence of the NOT/ALFOSC + Copernico/AFOSC and VLT/X-shooter host-subtracted spectra is shown in Figures \ref{fig:NOTspec_suball} and \ref{fig:XSHspec_suball}, respectively. In the spectra taken shortly after the light curve peak, very broad (FWHM$\sim$10$^{4}$ \kms)  emission lines from the H$\upbeta$, \ion{He}{II} $\lambda$4686 and H$\upalpha$ transitions, are clearly visible. Moreover, both in the early NOT/ALFOSC + Copernico/AFOSC and VLT/X-shooter spectra, we detected a strong broad emission line at the position of H$\updelta$. Furthermore, narrow [\ion{O}{III}] and [\ion{N}{II}] emission lines are detected only in the medium--resolution X-shooter spectra. While the broad components became narrower and gradually disappear with time, the [\ion{O}{III}] and [\ion{N}{II}] narrow emission lines are clearly visible also in the last VLT/X-shooter spectrum, suggesting that they are unrelated to the TDE emission, but instead they belong to the host galaxy spectrum. 

Interestingly, in the X-shooter spectra, the \ion{He}{II} $\lambda$4686 emission line  is double peaked and the separation between the two peaks becomes more prominent at later times. In previous work on iPTF16fnl, \cite{brown18} suggested the presence of a blue--shifted \ion{C}{III}/\ion{N}{III} blend in the blue wing of the \ion{He}{II}. Thanks to our medium--resolution X-shooter spectra we have been able to separate these two components. In particular, while the first feature is well centered on the \ion{He}{II} rest frame wavelength, the second has a blue--shifted best--fitting wavelength, compatible with the \ion{N}{III} $\lambda$4640. Furthermore, similarly to what found in the case of the TDE AT2018dyb by \cite{leloudas19}, the presence of an apparent H$\updelta$ emission line in our optical spectra is hard to explain as we do not detect the H$\upgamma$ emission line and both H$\upbeta$ and H$\upalpha$ are much fainter as well.  Instead, this strong emission line can be associated with the \ion{N}{III} $\lambda$4100 transition {\cite[][]{leloudas19}}, which is produced together with the \ion{N}{III} $\lambda$4640 in the the Bowen fluorescence mechanism \cite[]{bowen34, bowen35}.

The detection of Bowen emission lines have been already suggested in the interpretation of optical spectra of some TDEs. The presence of the \ion{N}{III}/\ion{C}{III} blend, usually detected in Wolf-Rayet stars and always seen in X-ray Binaries \cite[]{McClintock75} was discussed by \cite{gezari15} as a possible explanation of the blue wing observed in the \ion{He}{II} $\lambda$4686 in the optical spectra of the TDE PS1$-$10jh.  More recently, \cite{blagorodnova18} detected the \ion{O}{III} and \ion{N}{III} emission lines in the UV and optical spectra of the TDE iPTF15af and explained such transitions with the Bowen fluorescence mechanism. \cite{leloudas19} clearly detected Bowen fluorescence lines in the optical spectra of the TDE AT2018dyb. In particular, \cite{leloudas19}, when analyzing the optical spectra of past events available in literature, found that such lines are quite commons in TDEs. Thus the authors identify a N-rich subset among the tidal disruption events population. In a recent works, \cite{trakhtenbrot19} and \cite{gromadzki19} proposed a new class of nuclear transients associated with re-ignition of enhanced accretion on the SMBH  for AT2017bgt and OGLE17aaj.
They also found that F01004-2237, previously classified as TDE candidate by \citealt{tadhunter17}, belongs 
to this group. The optical spectra of these objects also shows Bowen fluorescence lines. 

Our medium--resolution observations of iPTF16fnl strongly indicate the presence of such features also in the optical spectra of this source and place this transient among the N-rich TDE subset. 


\subsection{Fits to the emission lines}
In order to investigate the properties and evolution of the main spectral features, we fitted the emission lines present in the host-subtracted spectra of iPTF16fnl shown in Figures \ref{fig:NOTspec_suball} and \ref{fig:XSHspec_suball}. 
We modelled the more prominent emission lines with Gaussian functions (but see also \citealt{roth18}) using the \texttt{python} packages \texttt{curvefit} and \texttt{leastsq}. In the case of broad emission lines, multi-components Gaussian fit have been applied, when needed. The emission features have been analyzed selecting a $\sim$ 1500 \AA\/ wide  fitting window which include both the broad features of interest and the local continuum. In the fitting procedure, both the central wavelength and FWHM of the lines have been left free to vary in the selected wavelength range. 
In the case of late-time spectra, where the width of broad emission lines became smaller and the narrow emission lines emerge, a narrower fitting window has been selected. 

\begin{figure*}
        \centering
        \begin{subfigure}[b]{0.475\textwidth}
            \centering
            \includegraphics[width=\textwidth]{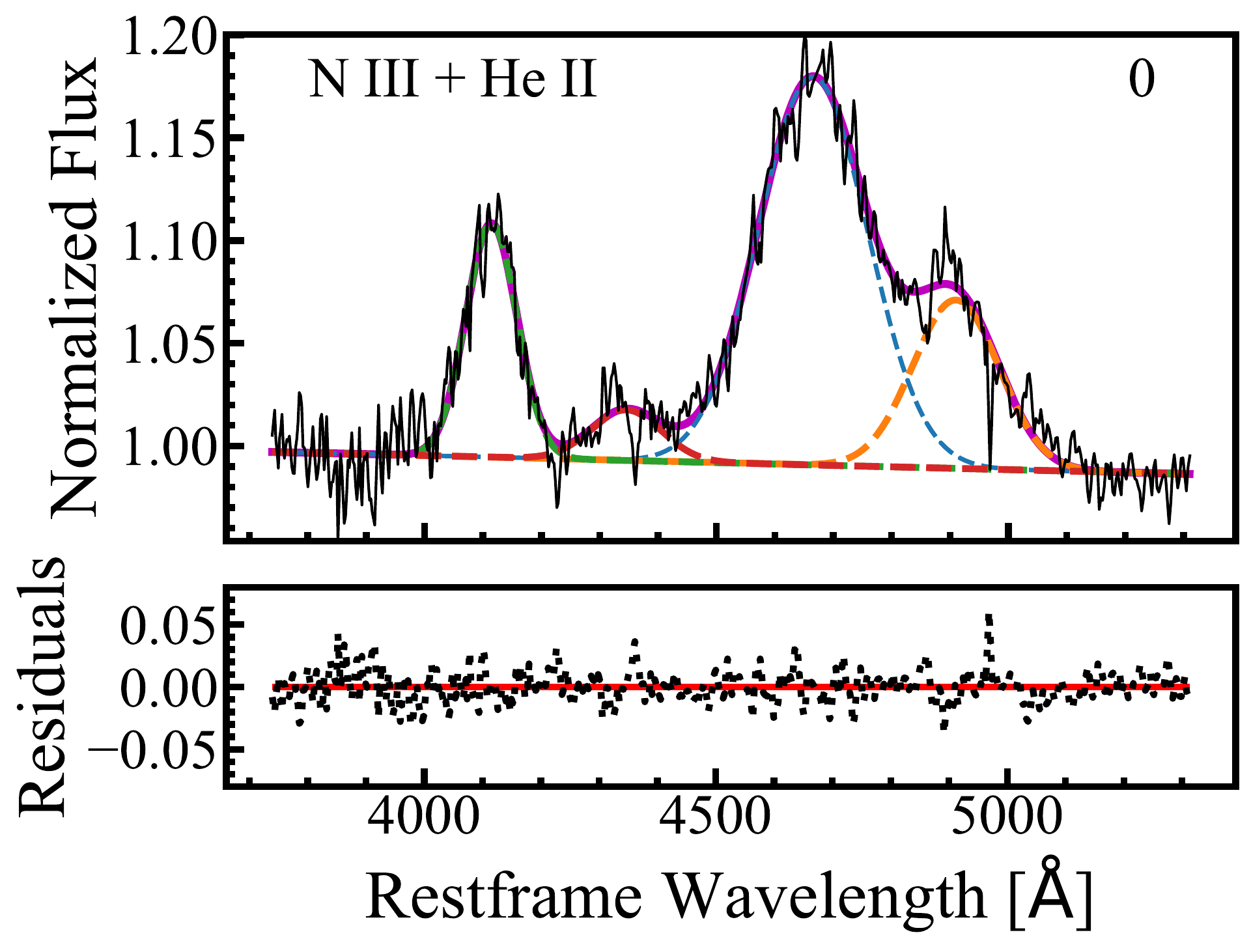}
        \end{subfigure}
        \hfill
        \begin{subfigure}[b]{0.475\textwidth}  
            \centering 
            \includegraphics[width=\textwidth]{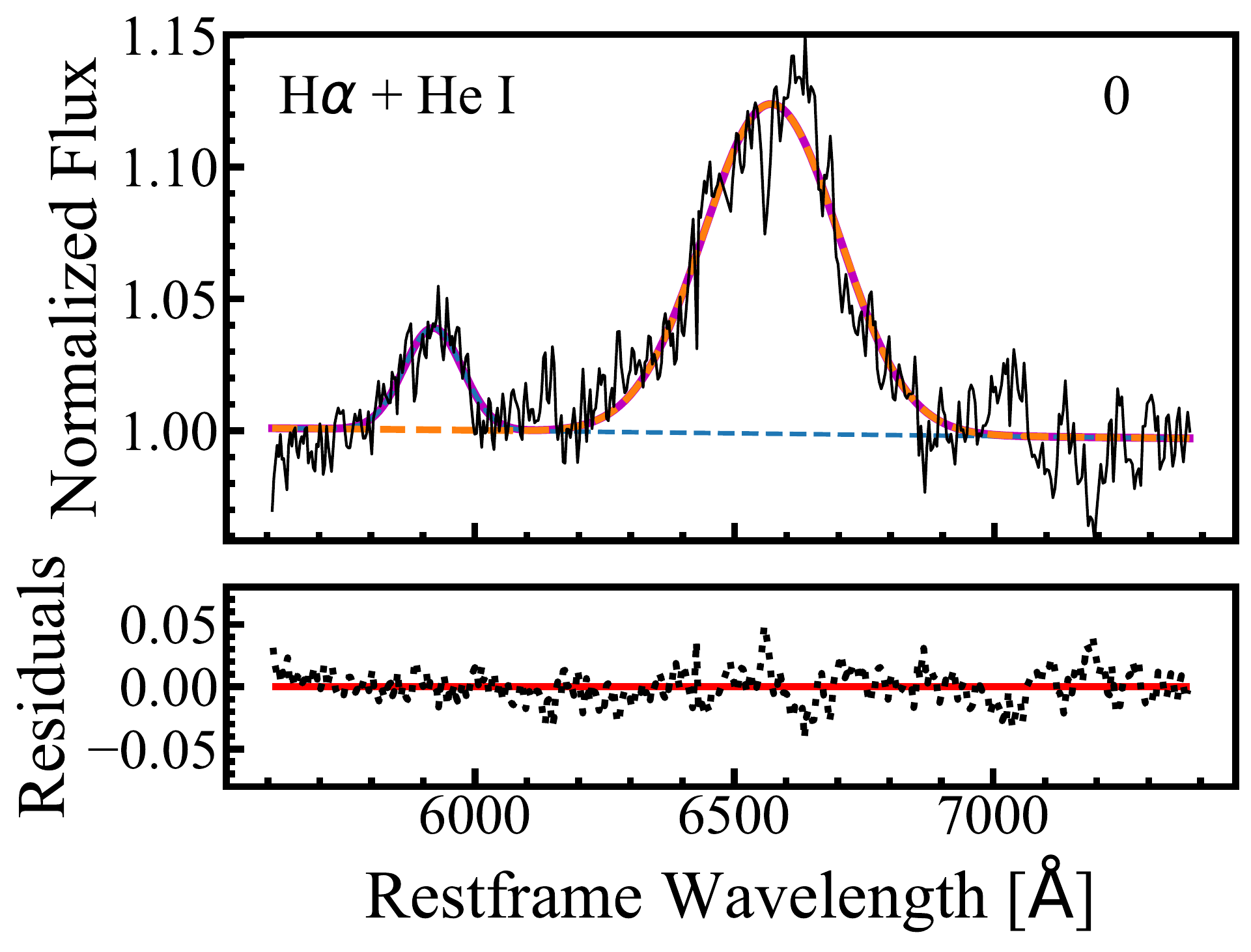}
        \end{subfigure}
        \vskip\baselineskip
        \begin{subfigure}[b]{0.475\textwidth}   
            \centering 
            \includegraphics[width=\textwidth]{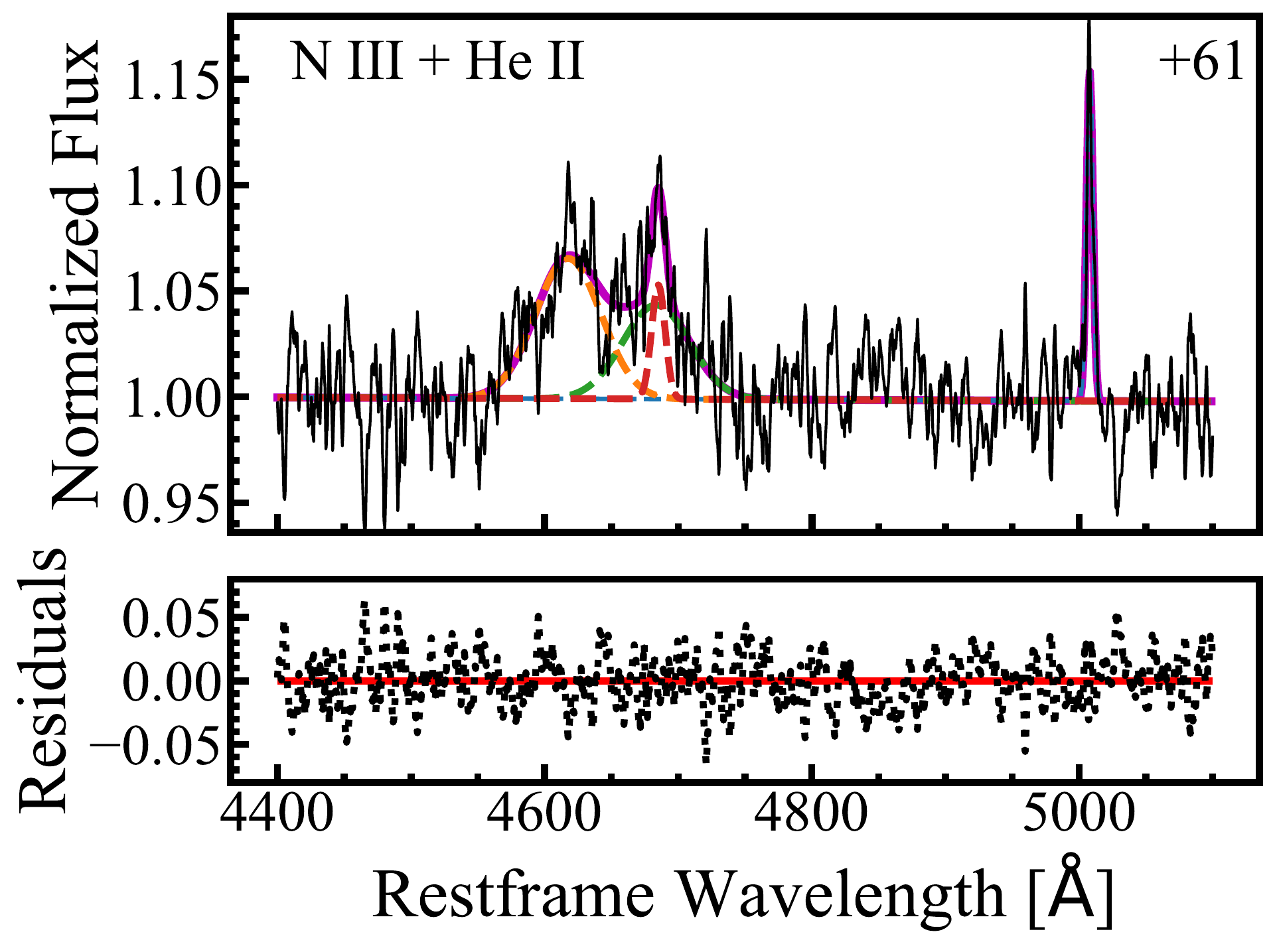}
        \end{subfigure}
        \quad
        \begin{subfigure}[b]{0.475\textwidth}  
            \centering 
            \includegraphics[width=\textwidth]{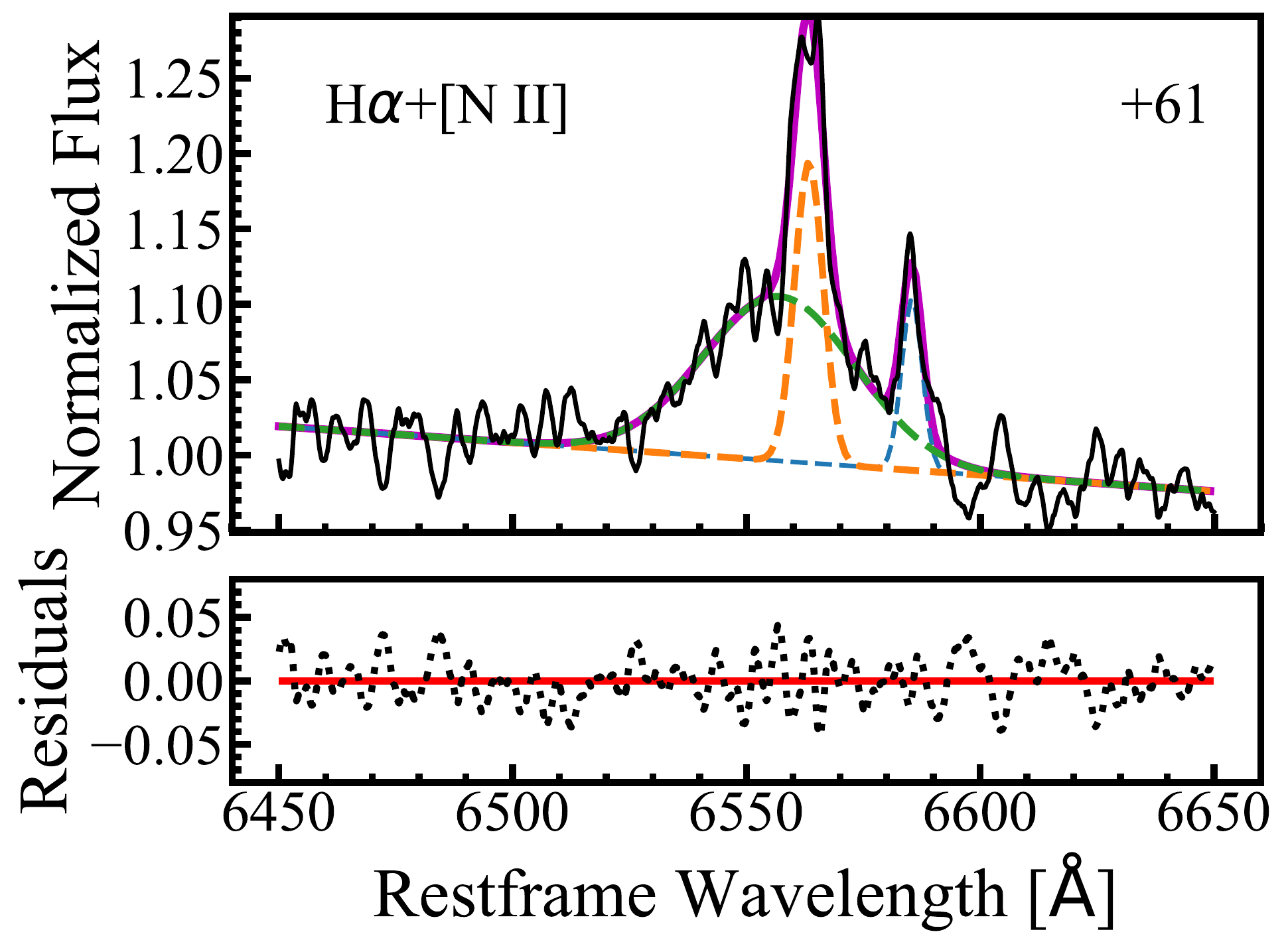}
        \end{subfigure}
        \caption{Multi-gaussian emission lines fit of the components found in the \ion{He}{II} and H$\upalpha$ regions in the host-subtracted spectra, taken at 0 and 61 days after the TDE light curve peak (upper and lower panels, respectively). The Gaussian components of each line are represented by colored dashed lines, while the magenta solid line show the total fitting model. Residuals are shown at the bottom of each panel.} 
        \label{fig:linefits}
    \end{figure*}

In Figure \ref{fig:linefits} the multi-component line fits together with the model residuals, performed in the \ion{He}{II} and H$\upalpha$ regions of the spectra taken at epochs 0 and 61 is shown. In the late-time epoch (lower panels of Figure \ref{fig:linefits}), narrow components emerge from the broad emission features. Such contribution has been accurately isolated through the multi-component fit. Although the development of a narrow core in the broad features at late-times is clearly visible, the Gaussian functions are well suited in modelling the emission line shapes, especially in the early-times. In the late-epochs medium resolution spectra the Gaussian multi-component fit is still a good approximation of the line profiles and it allows us to accurately separate the narrow line contribution from the whole emission line feature, in order to derive the main parameters of the broad line component (Figure \ref{fig:linefits}, lower panels).

\subsubsection{The NOT/ALFOSC and Copernico/AFOSC spectra}
In the NOT/ALFOSC and Copernico/AFOSC host subtracted spectra (Figure \ref{fig:NOTspec_suball}), strong broad emission lines and their evolution are clearly visible. In particular, the spectra taken soon after the light curve peak show strong and broad \ion{He}{II} $\lambda$4686 , \ion{N}{III} $\lambda$4100 and H$\upbeta$. The latter quickly became narrower and could no longer be detected in the late-time spectra. The \ion{N}{III} $\lambda$4100 is well separated from the broad \ion{He}{II} $\lambda$4686, thus it has been possible to model it with a single Gaussian component. Instead, the broad feature in the \ion{He}{II} wavelength region is more complex and, depending on the status of its evolution, it has been necessary to use more Gaussian components to accurately model it. For instance, in the first spectrum, we used two Gaussians in order to take into account also the contribution of the broad H$\beta$ emission, which blends at that epoch with the helium emission line (see Figure \ref{fig:linefits}, upper-right panel). In the spectrum taken two days after the light curve peak, the \ion{He}{II} $\lambda$4686 develops a strong blue wing which we ascribe to the \ion{N}{III} $\lambda$4640 emission line. Thus, we modelled this region using a total of three Gaussians, in order to take into account for the \ion{He}{II}, \ion{N}{III} and H$\upbeta$ contributions.

The broad H$\upalpha$ is clearly detected already in the first spectrum and it becomes narrower with time. The evolution is similar to that observed in the broad \ion{He}{II} $\lambda$4686. The emission line is well described by a single Gaussian function. 

Finally, we detected a broad emission line which we identify as \ion{He}{I} $\lambda$5875. This component become rapidly faint and is not detectable anymore already in the spectra taken $\sim$one month after the TDE peak (see Figures \ref{fig:NOTspec_suball} and \ref{fig:XSHspec_suball}). When detected, the \ion{He}{I} $\lambda$5875 emission line can be described well with a single Gaussian (see Figure \ref{fig:linefits}, upper-left panel).     

\subsubsection{The VLT/X-shooter optical spectra}
\label{subsubsec:xshopt}
The VLT/X-shooter host-subtracted optical spectra (Figure \ref{fig:XSHspec_suball}) are  dominated by broad \ion{He}{II} $\lambda$4686, \ion{N}{III} $\lambda$4100, H$\upbeta$ and H$\upalpha$ emission lines. The broad line profiles are quite symmetric, they do not shows signs of outflows and become more narrow and faint at later times. Moreover, narrow emission lines of [\ion{O}{III}]$\lambda\lambda$5007, 4959 and [\ion{N}{II}] $\lambda\lambda$6548,6584 are always detected, while narrow [\ion{S}{II}] emission lines are detected in late-time spectra. Only in two late-time spectra we identify a faint narrow H$\beta$ emission line. In Table \ref{tbl:nlineXS}, the main results of the narrow emission lines analysis are shown. While the more intense lines [\ion{O}{III}]$\lambda$5007, [\ion{N}{II}]$\lambda$6584 and H$\upalpha$, are always well detected, in some spectra it was not possible to fit the other narrow lines ([\ion{O}{III}]$\lambda$4959 and [\ion{N}{II}] $\lambda$6548) as they are too faint or in blend with broad components. 

\begin{figure}
\includegraphics[width=1\columnwidth]{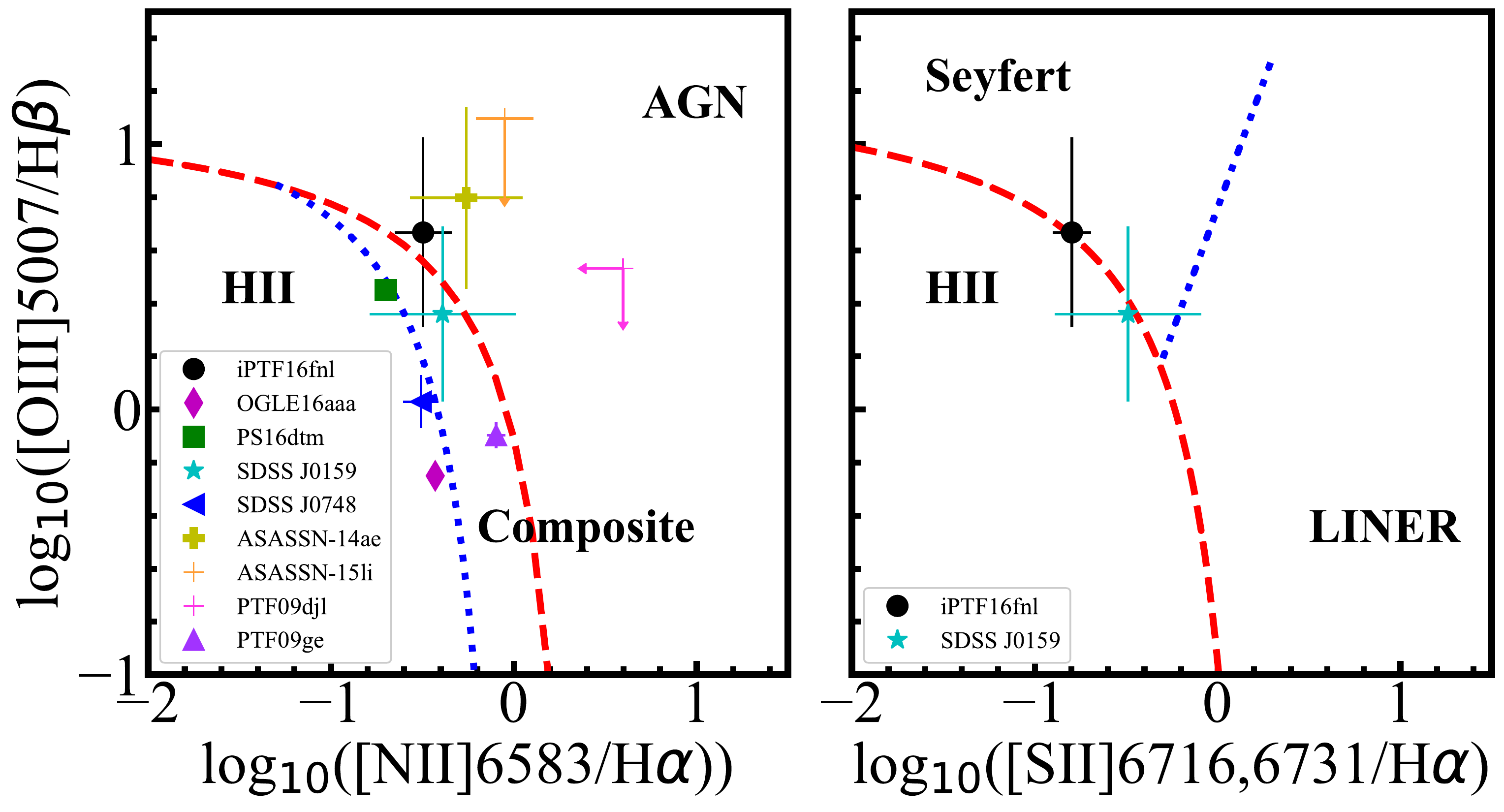}
\caption{Using the equivalent widths of the detected narrow emission lines we created BPT diagrams for the host galaxy of iPTF16fnl (black filled point). The lines separating the different activity regions are the following: red dashed line from \citep{kewley01} 
, blue dotted line from \citep{kauffmann03} in the left panel and blue dotted line from \citep{kewley06} in the right panel. For comparison we show the results for some TDEs for which line ratios are available in literature: OGLE16aaa (magenta filled diamond, \citealt{wyrzykowski17}), PS16dtm (green filled square, \citealt{blanchard17}), SDSS J0159+0033 (cyan filled star, \citealt{merloni15}), SDSS J0748 (filled blue  left-triangle, \citealt{french17}), ASASSN-14ae (filled yellow cross, \citealt{french17}), ASASSN-15li (orange plus, \citealt{french17}), PTF09djl (pink plus, \citealt{french17}) and PTF09ge (filled violet up-triangle, \citealt{french17}).   
}
\label{fig:BPT}
\end{figure}

The equivalent width and the FWHM of these narrow emission lines are consistent with being constant with time, regardless of the different seeing conditions. 
We thus ascribe this emission as coming from the host galaxy, and use this to investigate the properties of the host. The line ratios log([\ion{N}{II}]/H$\upalpha)$=($-$0.50$\pm$0.15), log([\ion{O}{III}]/H$\upbeta)$=(0.67$\pm$0.36) and log([\ion{S}{II}]/H$\upalpha)$=($-$0.80$\pm$0.10), derived from the latest X-shooter spectrum, suggest that the galaxy hosts a weak AGN in the core. In Figure \ref{fig:BPT} we show the location of iPTF16fnl line ratios in a Baldwin, Phillips \& Terlevich (BPT) diagram \citep{baldwin81}, together with the line ratios values for some TDE hosts found in galaxies with weak nuclear activity (OGLE16aaa,  \citealt{wyrzykowski17}; PS16dtm, \citealt{blanchard17}, SDSS J0159+0033, \citealt{merloni15} and SDSS J0748, ASASSN-14ae, ASASSN-15li, PTF09djl, PTF09ge from \citealt{french17}.  
Even if located at the boundary with the star--forming region, iPTF16fnl is among the TDEs with higher AGN activity signatures. 

The high-resolution \ion{He}{II} region is particularly interesting. Along with faint broad components of \ion{N}{III} $\lambda$4100 and H$\upgamma$, which are detected only in the first X-shooter spectrum, we clearly detect a double component in the broad \ion{He}{II} feature. The line is double-peaked and the presence of these two components become more evident with time. We ascribe the component blue-shifted with respect to the rest frame wavelength of \ion{He}{II} to the \ion{N}{III} $\lambda$4640 Bowen blend. Thus, we used two Gaussians in order to describe the broad feature; these Gaussians represent both the contribution of the \ion{He}{II} and the \ion{N}{III} emission lines (see Figure \ref{fig:linefits}, lower-left panel).

Along with the more prominent [\ion{O}{III}]$\lambda$5007 and $\lambda$4959, we identify a narrow \ion{He}{II} $\lambda$4686 emission line to be present alongside the broad features. Interestingly, this narrow feature appears only in late-time spectra, when the broad emission is less intense. This kind of line profile evolution, in which a narrow core is developed at later times along with the narrowing of the broad component, has been predicted to happen in the case of electron scattering broadening of the emission lines by \cite{roth18}. This suggest that the broad emission components observed in iPTF16fnl are emitted in an optically thick region in which the electron scattering play an important role in the broadening of the lines. As time pass by, the optical depth of this region decrease and the broad emission components develop a narrow core which become more prominent at later times.

\subsection{Evolution of the emission lines}
\begin{figure}
\includegraphics[width=1\columnwidth]{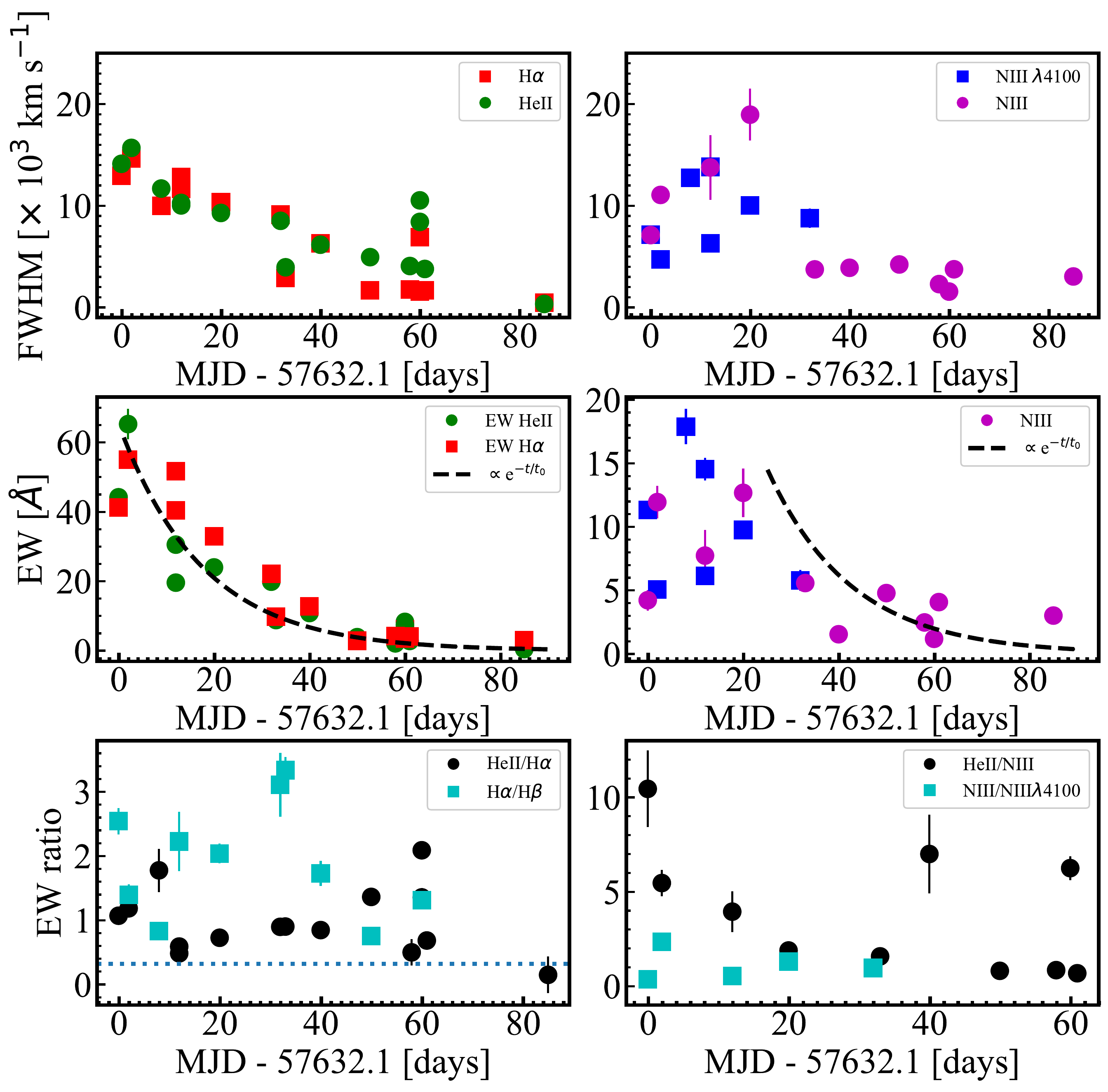}
\caption{ \textit {Upper-left}: time evolution of the FWHM of \ion{He}{II} and H$\upalpha$, green filled circles and red filled squares, respectively. \textit{Upper-right}: time evolution of the FWHM of \ion{N}{III} and \ion{N}{III}$\lambda$4100, magenta filled circles and blue filled squares, respectively. \textit{Center-left}: time evolution of the EW for \ion{He}{II} and H$\upalpha$ green filled circles and red filled squares, respectively. 
\textit {Center-right}: time evolution of the EW for \ion{N}{III} and \ion{N}{III}$\lambda$4100, magenta filled circles and blue filled squares, respectively. In both panels, the black dashed line indicate the e$^{-t/t_0}$ with t$_{0}$=17.6 evolution, found for the bolometric luminosity. \textit {Bottom-left}: time evolution of the line ratio \ion{He}{II}/H$\upalpha$ and H$\upalpha$/H$\upbeta$, black filled circles and cyan filled squares, respectively. \textit {Bottom-right}: time evolution of the line ratio \ion{He}{II}/\ion{N}{III} and \ion{N}{III}/\ion{N}{III}$\lambda$4100, black filled circles and cyan filled squares, respectively. The blue-dotted horizontal line indicate the values for Helium-to-Hydrogen ratio expected for a nebular environment and for solar abundance as reported by \citealt{hung17}.  
}
\label{fig:fitres}
\end{figure}

We studied the evolution with time of the main properties of the broad emission lines, inferred from the fits to the NOT/ALFOSC, Copernico/AFOSC  and VLT/X-shooter host-subtracted spectra. In Figures \ref{fig:fitres} we show the behaviour of the FWHM and the absolute values of the EW for \ion{He}{II}, H$\upalpha$, \ion{N}{III} lines (upper and central panels, respectively), along with the time evolution of the line ratios \ion{He}{II}/H$\upalpha$, H$\upalpha$/H$\upbeta$, \ion{He}{II}/\ion{N}{III} and \ion{N}{III}/\ion{N}{III}$\lambda$4100 (lower panels). 
All the reported FWHMs have been corrected for the instrumental broadening.
In general, we observe a reduction in the FWHM of all lines. The \ion{He}{II} and H$\upalpha$ show similar behaviour, starting from a FWHM$\sim$14$\times$10$^{3}$km s$^{-1}$ and reaching a value of FWHM$\sim$2$\times$10$^{3}$km s$^{-1}$ for the H$\upalpha$ and FWHM$\sim$4$\times$10$^{3}$kms$^{-1}$ for \ion{He}{II} 60 days after the light curve peak. In addition, both lines show a narrow component in the last medium--resolution X-shooter spectrum, with FWHM$\sim$500 km s$^{-1}$ and FWHM$\sim$300 km s$^{-1}$ for H$\upalpha$ and \ion{He}{II}, respectively. 
As already discussed in section \ref{subsubsec:xshopt}, we ascribe the narrow H$\upalpha$ to the host galaxy contribution. Instead, no narrow \ion{He}{II} component is detected along with the broad feature in the early-times optical spectra. A narrower component starts to emerge in the spectra $\sim$50 days after the TDE peak and the FWHM and EW decline over time.  Thus, we ascribe the narrow \ion{He}{II} line observed in the last X-shooter spectrum to the TDE reprocessing nebula emission, in which the drop in density over time produces the narrowing of the broad TDE features toward late phases. 

A similar behaviour is observed also in the FWHM of both the \ion{N}{III} emission lines, which declines with time, but only at later times, with the \ion{N}{III} $\lambda$4100 evolving more rapidly and already disappearing one month later. 
During the first 20 days from the TDE peak the FWHM of these lines show an increase with time, which is particularly evident in the \ion{N}{III} $\lambda$4640 behaviour. Indeed, this component is characterized by a FWHM=7$\pm$1 $\times$10$^{3}$ km s$^{-1}$ in the spectrum taken during the TDE light-curve peak and it rises until it reaches the value of FWHM=19$\pm$3 $\times$10$^{3}$ km s$^{-1}$ 20 days after. Afterwards, the width of this line follows the decline trend and it shows a broad component with FWHM$\sim$3$\times$10$^{3}$ km s$^{-1}$ in the last X-shooter spectrum. 

There is a clear evolution with time also in the EWs. While the EWs of \ion{He}{II} and H$\upalpha$ decline following a very similar behaviour, the trend for  \ion{N}{III} lines is more scattered in the early-time observations and become more clear after 20 days from the light curve peak. Interestingly, the EW time evolution observed in both \ion{He}{II} and H$\upalpha$ follow the exponential decline we found for the bolometric luminosity of iPTF16fnl. Also the EW of \ion{N}{III} follow a similar behaviour, but only at late times. This suggest that these lines are powered by the ionizing luminosity.   

The ratio of the equivalent widths of the \ion{He}{II}/H$\upalpha$ and \ion{N}{III}/\ion{N}{III}$\lambda$4100 lines are consistent with being constant with time. In particular, the equivalent width ratio \ion{He}{II}/H$\upalpha$ is always above the value expected for a nebular environment \cite[blue-dotted lines,][]{hung17}. Only in the observation taken 85 day after the TDE peak it approaches such value. In contrast, we observe a clear trend in the H$\upalpha$/H$\upbeta$ and \ion{He}{II}/\ion{N}{III} equivalent width ratios. The H$\upalpha$/H$\upbeta$ ratio passes from an initial value of 2.5$\pm$0.2 to a final value of 1.3$\pm$0.1 at day 60. Instead, the evolution of the ratio of the equivalent width of \ion{He}{II} to \ion{N}{III} is more dramatic and rapid, starting from an initial value of 10.5$\pm$2.0 to a final value of 0.7$\pm$0.1 at day 60, with an exponential decline.

\begin{figure*}
\includegraphics[width=2\columnwidth]{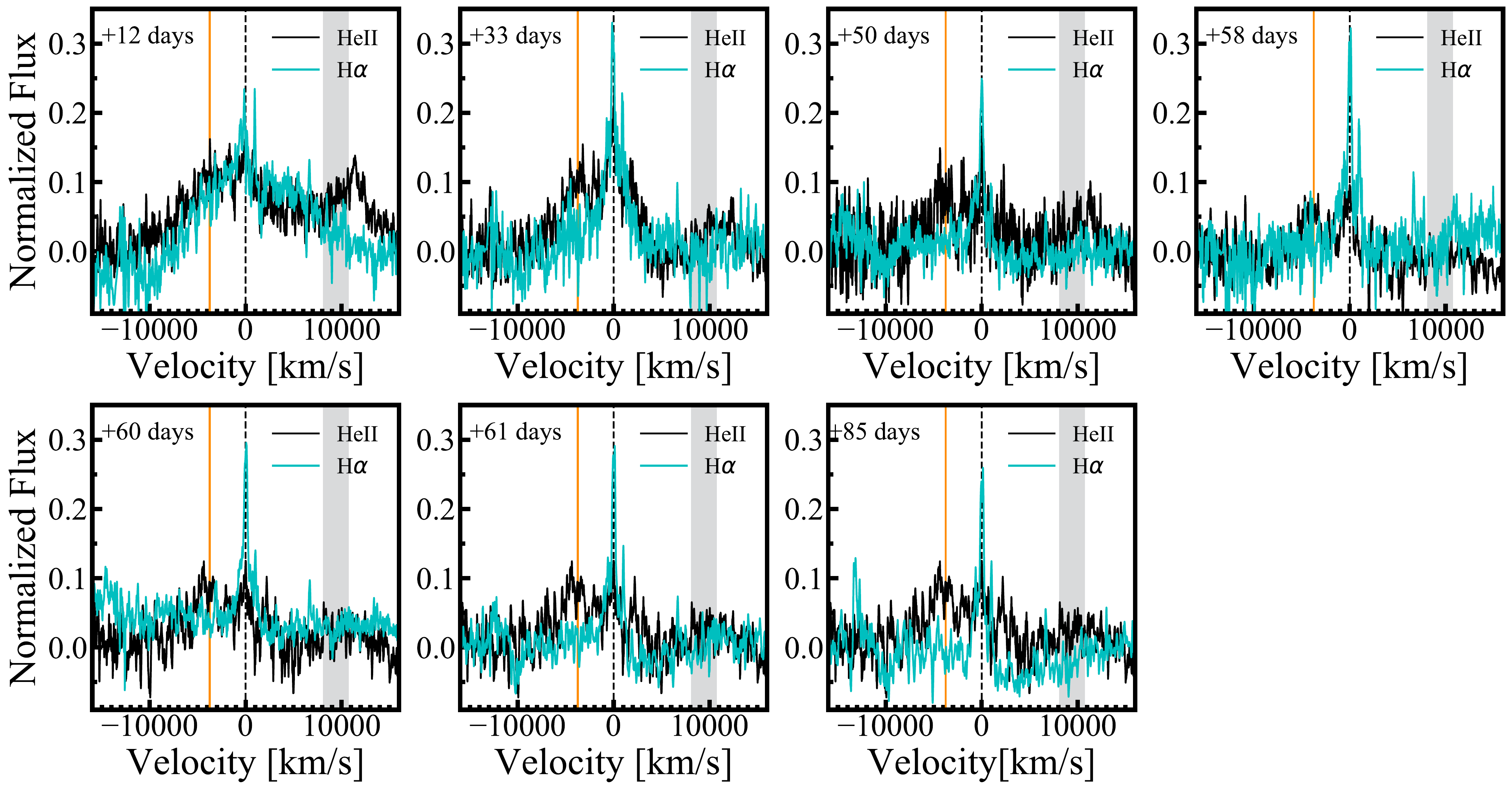}
\caption{Comparison between the \ion{He}{II} $\lambda$4686 (in black) and \ion{H}{$\alpha$} (in cyan) emission lines observed in the X-shooter spectra. Orange solid lines indicate the \ion{N}{III} $\lambda$4640 position. The gray bands indicate the H$_{\alpha}$ areas affected by telluric absorption. 
}
\label{fig:XSHvelocity}
\end{figure*}

In Figure \ref{fig:XSHvelocity} the comparison between the X-shooter \ion{He}{II} and H$\upalpha$ emission line profile as a function of time is shown. The line profile evolution is consistent with being the same for both lines. The narrowing of the broad components as well as the developing of the narrow core is clearly visible. 
The FWHM evolution of  H$\upalpha$ and \ion{He}{II} is faster than what observed for the FWHM of \ion{N}{III} line. Indeed, already after 33 days from the light curve peak, the \ion{N}{III} component is well detected. As the \ion{He}{II} broad component become more narrow, following a trend similar to the broad H$\upalpha$, the \ion{N}{III} component become more prominent. Finally, in the last epoch, while a broad \ion{N}{III} component (FWHM =3046$\pm$166 km s$^{-1}$) is still detected, only narrow emission lines for \ion{He}{II} and H$\upalpha$ are visible.  

Also the shape of the \ion{He}{II} and H$\upalpha$ line profiles appears to be very similar. They both become narrower with time and in the last epoch spectrum only narrow components (FWHM =340$\pm$43 km s$^{-1}$ and FWHM = 487$\pm$9 km s$^{-1}$ for \ion{He}{II} and H$\upalpha$, respectively) are detected. Moreover, starting from the spectra taken 33 days after the light curve peak, when the more prominent broad components become fainter, a broad absorption on the red wing of both \ion{He}{II} and H$\alpha$ line profiles is visible.

\section{Discussion and Conclusions}

Our follow-up campaign of the tidal disruption event iPTF16fnl covers $\sim$100 days of the source emission and includes high--quality optical photometry as well as regular spectroscopic observations. The latter includes medium--resolution X-shooter spectroscopy. We have included data from UVOT/\textit{Swift}, which monitored the source for $\sim$ 300 days in our analysis.

Our bolometric light curve confirm that iPTF16fnl is a fast and faint TDE, as found by \cite{blagorodnova17}. Indeed, we have found a peak value of the bolometric luminosity of $L_{\rm p} \sim$4$\times$10$^{42}$ \unitlum, an order of magnitude fainter than what is usually found in TDEs. Its time evolution follows an exponential decline rather than the standard t$^{-5/3}$ power law and it is characterized by an e-folding time of $\sim$17 days, which is remarkably fast. 

From the black body fit we have derived the BB temperature and radius as well as their evolution with time. While $T_{\rm BB}$ is consistent with being $\sim$1.5$\times$10$^{4}$ K at all phases, there is a clear evolution in $R_{\rm BB}$. In particular, the BB radius expands during the epochs prior to the luminosity peak, when it reaches its maximum values of $R_{\rm BB}$=(3.6$\pm$0.3)$\times$10$^{14}$cm. Afterwards, it follows a declining trend until it reaches its minimum value of $R_{\rm BB}$=(0.8$\pm$0.4)$\times$10$^{14}$cm.
We note that this value is $\sim$three order of magnitudes larger than the Schwartzchild radius expected for a black hole with mass $M_{\rm BH} \sim$3$\times$10$^{5}$\Msun, measured for iPTF16fnl by \cite{wevers17}. When compared with the tidal radius expected for the disruption of a solar-like star, such value is still $\sim$one order of magnitude larger. In recent works,  similar results have been found on a sample of known TDEs \cite[]{hung17, wevers17, wevers19}, suggesting that, in these cases, the optical emission can be explain either with stream self-intersection models, or as produced in a reprocessing envelope at large radii from the SMBH, which shrinks over time. 
Our results from the photometric and the spectroscopic analysis suggest that the reprocessing envelope model can be applied to the optical emission of iPTF16fnl.  

Indeed, further indications of the presence of a reprocessing envelope are obtained through the spectroscopic analysis, both from emission lines identifications and broad components time evolution.    

We clearly detect strong \ion{N}{III} $\lambda$4100 and \ion{N}{III} $\lambda$4640 emission lines in the host-subtracted optical spectra. 
The \ion{N}{III} $\lambda$4100 transition is particularly intense in the early-time NOT/ALFOSC spectra and rapidly become fainter with time. The associated \ion{N}{III} $\lambda$4640 blends with the broad \ion{He}{II} line in the NOT/ALFOSC spectra but it is evident in the higher resolution X-shooter spectra, as a second peak close to the wavelength of the \ion{He}{II}. 

The detection of such transitions place iPTF16fnl among the newly discovered N-rich TDE subset, identified by \cite[]{leloudas19}.

These \ion{N}{III} transitions are known to be produced in the Bowen fluorescence mechanism together with a series of optical lines such as \ion{O}{III} at $\lambda\lambda$3047, 3133, 3312, 3341, 3444, 3760 and \ion{N}{III} at $\lambda\lambda$ 4097, 4379, 4634 \cite[]{osterbrock74} and are primarly triggered by the \ion{He}{II} ionization. While we clearly detect the \ion{N}{III} Bowen lines in iPTF16fnl spectra, there is no sign of the \ion{O}{III} lines. 

The early-phases observations of iPTF16fnl are performed mainly with NOT/ALFOSC instrument and the wavelength region where the Bowen \ion{O}{III} lines are expected is out of the instrumental spectral range. 
Instead, the X-shooter UVB arm include the wavelength range of interest, 
However, these observations started at later phases and a fast evolution in the \ion{O}{III} lines could explain their non detection in the X-shooter spectra. Indeed, a similar trend has been observed in the TDE AT2018dyb, where a faint \ion{O}{III} $\lambda$3760 has been detected in the early-phases spectra and it disappeared after $\sim$30 days, while the \ion{N}{III} components are still clearly detected after $\sim$90 days \cite[][]{leloudas19}.  



The Bowen fluorescence mechanism requires the emission of a large flux of extreme UV (EUV) photons to excite the involved ions and high optical depths in order to work efficiently ($\tau \sim$ 10$^3$ for an electron density of $N_{\rm e} \sim$ 10$^{6-9}$ cm$^{-3}$ in the case of symbiotic stars \cite[]{selvelli07, Hyung18} and $\tau \sim$ 10$^{3-6}$ for an hydrogen density of  $N_{\rm H} \sim$ 10$^{9.5}$ cm$^{-3}$ in the case of AGNs \cite[]{netzer85}). 
Similarly in what found for AT2018dyb \cite[]{leloudas19}, the detection of such transitions in the spectra of iPTF16fnl strongly suggests that the broad emission lines are emitted in an optically thick nebula, where the high densities favour  the occurrence of multiple scatterings, needed for an efficient Bowen fluorescence mechanism.

We observe that the FWHM of the \ion{He}{II}, H$\alpha$ and \ion{N}{III} broad components decline with time. The trend is nearly the same for the \ion{He}{II} and H$\alpha$  lines. We also found that a narrow emission line close to the rest--wavelength appears on top of the broad components at later times. These narrow lines become more prominent at later times. 

The shape of the lines profile together with the narrowing with time of the broad features are in agreement with the prediction of \cite{roth18} for the electron scattering line broadening in the case of high optical dephts emitting regions. In this scenario, the observed decrease of the broad emission line width in iPTF16fnl can be explained in terms of a decrease in the optical depth of the line emitting region with time.

Interestingly, although the presence of high ionization emission lines is indicative of the production of strong EUV or X-ray photons from the ionizing source, no X-ray emission has been detected for iPTF16fnl. Recently, \cite{leloudas19} found that in most of the optically selected TDE in which Bowens lines have been detected, no X-ray emission has been observed. Obscuration effects due the presence of an optically thick envelope could be responsible for the lack of X-ray emission in such systems. Furthermore, \cite{dai18} proposed an unified model for TDEs in which an electron scattering photosphere is present along the accretion disk but it is truncated near the poles of the system. In this scenario, the detection of the X-ray emission depends on the viewing angle. The lack of X-ray emission for iPTF16fnl can be explained within this model implying a relatively high inclination angle ($\gtrsim 68 \deg$ in the case of high inclination escaping spectrum, shown in \cite{dai18}, Figure 5).  

The EW evolution of \ion{He}{II}, H$\upalpha$ and \ion{N}{III} $\lambda$4640 (but only after $\sim$20 days from the light curve peak) follows the same exponential decline found for the bolometric luminosity, suggesting that this lines are powered by the ionizing luminosity.

We observe a clear evolution with time in the ratios of the equivalent width for H$\upalpha$/H$\beta$ and \ion{He}{II}/\ion{N}{III}, which follows, in both cases, a declining trend. The value for the Balmer lines ratio is close to three for observations near the TDE peak. This is the value expected for the ratio H$\upalpha$/H$\beta$ in case B recombination for zero extinction. However it continuously changes over time until it reaches values close to one $\sim$60 days after the light curve peak. 

Instead, the \ion{He}{II}/\ion{N}{III} ratio shows a more dramatic decline, starting from a value close to ten at the TDE light curve peak and reaching a value close to one 60 days after. 

The evolution in the line ratio reflects a variation over time in the physical condition in the TDE emitting region (i.e. the ionizing flux, density, optical depth). However, the development of models for radiative transfer and of diagnostic tools involving these transitions is needed in order to investigate on the TDE emitting region properties and its time evolution. \cite[][]{netzer85, leloudas19}.

Finally, we used the host narrow emission lines detected in the last X-shooter spectrum to study the properties of the iPTF16fnl host galaxy and we have found indications that the galaxy hosts a weak AGN in the nucleus.

\section*{Acknowledgements}

{We thank the referee for the useful comments that improved the manuscript}. FO, GC, PGJ and ZKR acknowledge support from European Research Council Consolidator Grant 647208. FO acknowledge the support of the H2020 Hemera program, grant agreement No 730970. MF is supported by a Royal Society - Science Foundation Ireland University Research Fellowship. M.S. is supported by a generous grant (13261) from VILLUM FONDEN and a project grant from the IRFD (Independent Research Fund Denmark).  JH acknowledges financial support from the Finnish Cultural Foundation. TW is funded in part by European Research Council grant 320360 and by European Commission grant 730980. MG is supported by the Polish NCN MAESTRO grant 2014/14/A/ST9/00121. N.E.-R. acknowledges support from the Spanish MICINN grant ESP2017-82674-R and FEDER funds. SB is partially supported by the PRIN- INAF 2016 with the project “Toward the SKA and CTA era: discovery, localisation, and physics of transient sources. NUTS is supported in part by IDA (The Instrument Centre for Danish Astronomy)." The data presented here were obtained [in part] with ALFOSC, which is provided by the Instituto de Astrofisica de Andalucia (IAA) under a joint agreement with the University of Copenhagen and NOTSA. Based [in part] on observations made with the Nordic Optical Telescope, operated by the Nordic Optical Telescope Scientific Association at the Observatorio del Roque de los Muchachos, La Palma, Spain, of the Instituto de Astrofisica de Canarias. The data presented here include observations collected at Copernico telescope (Asiago, Italy) of the INAF - Osservatorio Astronomico di Padova. The research leading to these results has received funding from the European Union's Horizon 2020 Programme under the AHEAD project (grant agreement n. 654215).




\bibliographystyle{mnras}
\bibliography{mybib} 

\appendix

\section{Photometry data}

\begin{table*}
\centering
\begin{minipage}{150mm}
 \caption{NOT/ALFOSC photometric measurements}
 \label{tbl:phot}
 \begin{center}
 \begin{tabular}{@{}llccccccc}
 \hline
MJD      &  Phase & \textit {u$^{\prime}$}          &  \textit{B}             & \textit {V}              & \textit {g$^{\prime}$}              & \textit {r$^{\prime}$}              & \textit {i$^{\prime}$}          & \textit {z$^{\prime}$} \\
(1)      & (2)            & (3)            &(4)             & (5)            & (6)            & (7)        & (8)    &(9) \\
\hline
57\,631.97 & 0  &17.29$\pm$0.07 & 17.22$\pm$0.08 & 17.21$\pm$0.07 & 17.08$\pm$0.07 & 17.37$\pm$0.05 & 17.66$\pm$0.04 &17.80$\pm$0.04\\
57\,634.10 & 2  &   $\cdots$        & 17.17$\pm$0.07 & 17.26$\pm$0.08 &  $\cdots$ &       $\cdots$      &        $\cdots$    &         $\cdots$  \\
57\,644.14 &12  &17.99$\pm$0.07 & 17.84$\pm$0.06 & 17.84$\pm$0.20 & 17.80$\pm$0.10 & 18.06$\pm$0.04 & 18.23$\pm$0.04 &18.40$\pm$0.03\\
57\,652.04 &20  &18.52$\pm$0.07 & 18.43$\pm$0.07 & 18.60$\pm$0.10 & 18.33$\pm$0.09 & 18.35$\pm$0.03 & 18.94$\pm$0.05 &18.94$\pm$0.08\\
57\,664.12 &32  &19.30$\pm$0.40 & 19.14$\pm$0.09 & 19.50$\pm$0.10 & 19.00$\pm$0.10 & 19.35$\pm$0.05 & 19.62$\pm$0.06 &19.60$\pm$0.20\\
57\,672.12 &40  &20.11$\pm$0.06 & 19.83$\pm$0.30 & 20.70$\pm$0.40 & 19.50$\pm$0.20 & 19.98$\pm$0.03 & 19.91$\pm$0.05 &20.40$\pm$0.20\\
57\,679.11 &47  &19.90$\pm$0.30 & 20.70$\pm$0.20 & 20.2$^{\ddagger}$ & 20.00$\pm$0.20 & 20.1$^{\ddagger}$&21.3$^{\ddagger}$&20.7$^{\ddagger}$\\
57\,692.05 &60  &20.70$^{\ddagger}$&20.3$^{\ddagger}$& 21.2$^{\ddagger}$& 20.60$\pm$0.20&20.3$^{\ddagger}$&21.4$^{\ddagger}$&20.8$^{\ddagger}$\\
\hline
\end{tabular}
\end{center}
Notes: (1) MJD date of observations; (2) Phase (days) with respect to the estimated date of TDE peak MJD 57\,632.1 according to \cite{blagorodnova17}, (3), (6), (7), (8) and (9) host-subtracted apparent magnitudes and uncertainties in the Sloan filters \textit {u$^{\prime}$},\textit { g$^{\prime}$}, \textit {r$^{\prime}$}, \textit {i$^{\prime}$} and \textit {z$^{\prime}$}, respectively, in the AB system; (4) and (5) host-subtracted apparent magnitudes and uncertainties in the Johnson filters \textit{B} and \textit{V}, respectively, in Vega system. The values indicated with $\ddagger$ are the 3$\sigma$ upper limits.
All the magnitudes reported are uncorrected for foreground extinction.
With  $\cdots$ we indicate epochs with no data available (no observations).
\noindent
\end{minipage}
\end{table*} 

\begin{table*}
\centering
 \begin{minipage}{140mm}
 \caption{ UVOT/\textit{ Swift} photometric measurements (AB system)}
 \label{tbl:UVphot}
 \begin{center}
 \begin{tabular}{@{}lccccccc}
 \hline
 MJD      & phase & \multicolumn{2}{c}{\textit {UVW2}}  & \multicolumn{2}{c}{\textit {UVM2}}& \multicolumn{2}{c}{\textit {UVW1}}  \\    
          & days  & mag & F$_{\lambda}$  & mag & F$_{\lambda}$  &  mag & F$_{\lambda}$   \\
(1)       & (2)   & (3)  &  (4) & (5) & (6) & (7)& (8)  \\
\hline
57\,630.82  & $-$1.28  & 16.49$\pm$0.04 & 66.8$\pm$2.30 & 16.81$\pm$0.05& 41.6$\pm$1.4 & 16.73$\pm$0.05& 32.7$\pm$1.6\\
57\,635.25  & 3.15 & 16.73$\pm$0.04 & 53.3$\pm$2.00 & 16.98$\pm$0.04& 35.6$\pm$0.9 & 16.84$\pm$0.07& 29.3$\pm$1.3\\
57\,636.59  & 4.49 & 16.81$\pm$0.03 & 49.6$\pm$1.60 & 17.06$\pm$0.05& 33.1$\pm$1.2 & 16.90$\pm$0.05& 27.9$\pm$1.3\\
57\,638.31  & 6.21  & 16.93$\pm$0.05 & 44.7$\pm$1.10 & 17.15$\pm$0.05& 30.4$\pm$1.1 & 17.07$\pm$0.05& 23.8$\pm$1.3\\
57\,639.45  & 7.35 & 17.01$\pm$0.04 & 41.5$\pm$1.50 & 17.25$\pm$0.05& 27.6$\pm$1.0 & 17.05$\pm$0.06& 24.2$\pm$1.3\\
57\,640.12  & 8.02 & 16.99$\pm$0.04 & 42.0$\pm$1.50 & 17.26$\pm$0.06& 27.5$\pm$1.1 & 17.20$\pm$0.07& 21.0$\pm$1.3 \\
57\,642.57  & 10.47  &  17.18$\pm$0.05 & 35.3$\pm$1.60  & 17.43$\pm$0.06& 23.5$\pm$1.0 & 17.35$\pm$0.07& 18.4$\pm$1.2 \\
57\,643.57  & 11.47 & 17.19$\pm$0.04 & 35.2$\pm$1.20 & 17.58$\pm$0.05& 20.5$\pm$0.8 & 17.39$\pm$0.06& 17.7$\pm$0.9 \\
57\,645.82  & 13.72  & 17.41$\pm$0.04 & 28.7$\pm$1.10 & 17.66$\pm$0.06& 18.9$\pm$0.8 & 17.49$\pm$0.07& 16.1$\pm$0.9 \\
57\,648.69  & 16.59 & 17.64$\pm$0.05 & 23.2$\pm$0.90 & 17.83$\pm$0.07& 16.2$\pm$0.8 & 17.68$\pm$0.08& 13.5$\pm$1.0 \\
57\,651.40  & 19.30  & 17.80$\pm$0.05 & 19.9$\pm$1.60 & 18.04$\pm$0.06& 13.4$\pm$0.6 & 17.84$\pm$0.07& 11.7$\pm$0.8 \\
57\,655.39  & 23.29 & 18.00$\pm$0.05 & 16.6$\pm$0.80 & 18.27$\pm$0.07& 10.8$\pm$0.6 & 18.03$\pm$0.08&  9.8$\pm$0.7 \\
57\,655.45  & 23.35 & 17.96$\pm$0.08 & 17.3$\pm$1.10 & 18.27$\pm$0.09& 10.8$\pm$0.7 & 17.91$\pm$0.09& 11.0$\pm$0.9 \\
57\,657.51  & 25.41 & 18.25$\pm$0.06 & 13.2$\pm$0.70 & 18.40$\pm$0.06&  9.6$\pm$0.5 & 18.10$\pm$0.07&  9.2$\pm$0.6 \\
57\,660.37  & 28.27 & 18.31$\pm$0.05 & 12.5$\pm$0.60 & 18.47$\pm$0.07&  9.0$\pm$0.5 & 18.05$\pm$0.07&  9.7$\pm$0.6 \\
57\,663.63  & 31.53  & 18.45$\pm$0.03 & 11.0$\pm$0.60 & 18.48$\pm$0.07&  9.0$\pm$0.5 & 18.27$\pm$0.09&  7.9$\pm$0.6  \\
57\,667.55  & 35.45  & 18.61$\pm$0.06 & 9.51$\pm$0.50 & 18.71$\pm$0.08&  7.2$\pm$0.5 & 18.24$\pm$0.09&  8.1$\pm$0.6  \\
57\,671.60  & 39.50  & 18.75$\pm$0.08 & 8.33$\pm$0.59 & $\cdots$&   $\cdots$ & 18.38$\pm$0.09&  7.1$\pm$0.6  \\
57\,676.13  & 44.03 & 18.72$\pm$0.10 & 8.54$\pm$0.77 & 18.80$\pm$0.13&  6.6$\pm$0.7 & 18.49$\pm$0.12&  6.4$\pm$0.7  \\
57\,679.57  & 47.47 & 18.94$\pm$0.06 & 6.97$\pm$0.41 & 18.87$\pm$0.07&  6.2$\pm$0.4 & 18.48$\pm$0.08&  6.5$\pm$0.5  \\
57\,683.09  & 50.99  & 18.84$\pm$0.07 & 7.67$\pm$0.46 & 18.94$\pm$0.07&  5.8$\pm$0.4 & 18.50$\pm$0.08&  6.4$\pm$0.5  \\ 
57\,687.82  & 55.72 & 19.06$\pm$0.08 & 6.25$\pm$0.44 & 19.03$\pm$0.09&  5.4$\pm$0.4 & 18.66$\pm$0.09&  5.5$\pm$0.5  \\
57\,691.86  & 59.76 & 19.05$\pm$0.08 & 6.33$\pm$0.41 & 19.07$\pm$0.08&  5.2$\pm$0.3 & 18.57$\pm$0.08&  6.0$\pm$0.4  \\
57\,712.74  & 80.64 & 19.23$\pm$0.09 & 5.37$\pm$0.45 & 19.41$\pm$0.15&  3.8$\pm$0.5 & 18.61$\pm$0.10&  5.7$\pm$0.5  \\
57\,716.72  & 84.62 & 19.28$\pm$0.06 & 5.11$\pm$0.29 & 19.30$\pm$0.10&  4.2$\pm$0.3 & 18.87$\pm$0.09&  4.5$\pm$0.3  \\
57\,720.04  & 87.94 & 19.45$\pm$0.10 & 4.37$\pm$0.42 & 19.42$\pm$0.16&  3.8$\pm$0.5 & 18.75$\pm$0.11&  5.1$\pm$0.6  \\
57\,724.71  & 92.61 & 19.41$\pm$0.07 & 4.54$\pm$0.27 & 19.07$\pm$0.18&  5.2$\pm$0.8 & 18.89$\pm$0.16&  4.5$\pm$0.7  \\
57\,728.09  & 95.99 & 19.38$\pm$0.08 & 4.64$\pm$0.34 & 19.43$\pm$0.13&  3.7$\pm$0.4 & 18.73$\pm$0.09&  5.1$\pm$0.5  \\
57\,732.40  & 100.30 & 19.34$\pm$0.07 & 4.85$\pm$0.27 & 19.27$\pm$0.11&  4.3$\pm$0.4 & 18.92$\pm$0.09&  4.3$\pm$0.4  \\
57\,740.63  & 108.53  & 19.32$\pm$0.11 & 4.93$\pm$0.52 & 19.44$\pm$0.11&  3.7$\pm$0.4 & 18.78$\pm$0.09&  4.9$\pm$0.4  \\
57\,743.26  & 111.16  & 19.39$\pm$0.08 & 4.61$\pm$0.33 & 19.44$\pm$0.12&  3.7$\pm$0.4 & 18.68$\pm$0.09&  5.4$\pm$0.4 \\
57\,744.50  & 112.40  & 19.54$\pm$0.10 & 4.03$\pm$0.37 & 19.29$\pm$0.13&  4.2$\pm$0.5 & 18.88$\pm$0.10&  4.5$\pm$0.4 \\
57\,748.55  & 116.45  & 19.44$\pm$0.08 & 4.40$\pm$0.29 & 19.32$\pm$0.13&  4.1$\pm$0.5 & 18.77$\pm$0.10&  5.0$\pm$0.5 \\
57\,932.05  & 299.95  & 19.47$\pm$0.11 & 4.28$\pm$0.45 & 19.96$\pm$0.15&  2.3$\pm$0.3 &  18.92$\pm$0.12&  4.3$\pm$0.5   \\
 \hline  
 \end{tabular}
 \end{center}
Notes: (1) MJD date of observations; (2) Phase (days) with respect to the estimated date of TDE peak MJD 57\,632.1 according to \citealt{blagorodnova17}; (3) \textit {UVW2} apparent magnitude and uncertainties; (4) \textit{ UVW2} flux density and uncertainties; (5) \textit{ UVM2} apparent magnitude and uncertainties; (6) \textit {UVM2} flux density and uncertainties; (7) \textit {UVW1} apparent magnitude and uncertainties; (8) \textit{ UVW1} flux density and uncertainties. Flux densities are in $\times$10$^{-16}$ [erg s$^{-1}$ cm$^{-2}$ \AA$^{-1}$].\\
 \noindent
 \end{minipage}
 \end{table*} 
 
 \section{Spectroscopic data}
 
 \begin{figure*}
\includegraphics[scale=0.7, angle = 0]{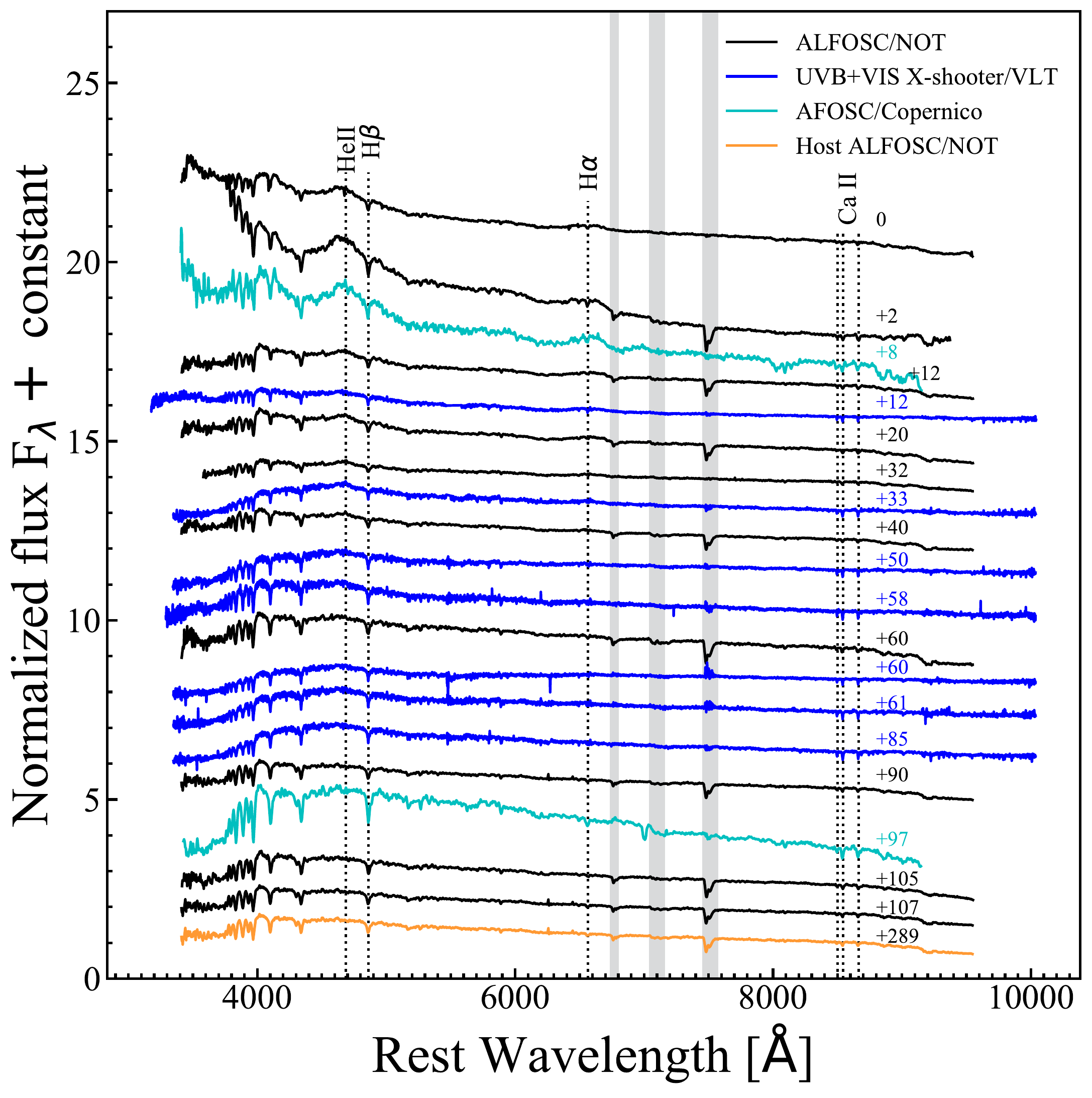}
\caption{Sequence of the rest-frame optical spectra of iPTF16fnl taken with NOT/ALFOSC (in black), AFOSC/Asiago (in cyan) and X-shooter (in blue). All the spectra have been corrected for reddening. The time of the observation in days since the time of the peak of the light curve and the main emission lines are indicated. The location of telluric absorption lines is indicated by gray bands. The NOT/ALFOSC host galaxy spectrum is shown in orange.
}
\label{fig:NOTspec}
\end{figure*}

\begin{figure*}
\includegraphics[scale=0.7, angle = 0]{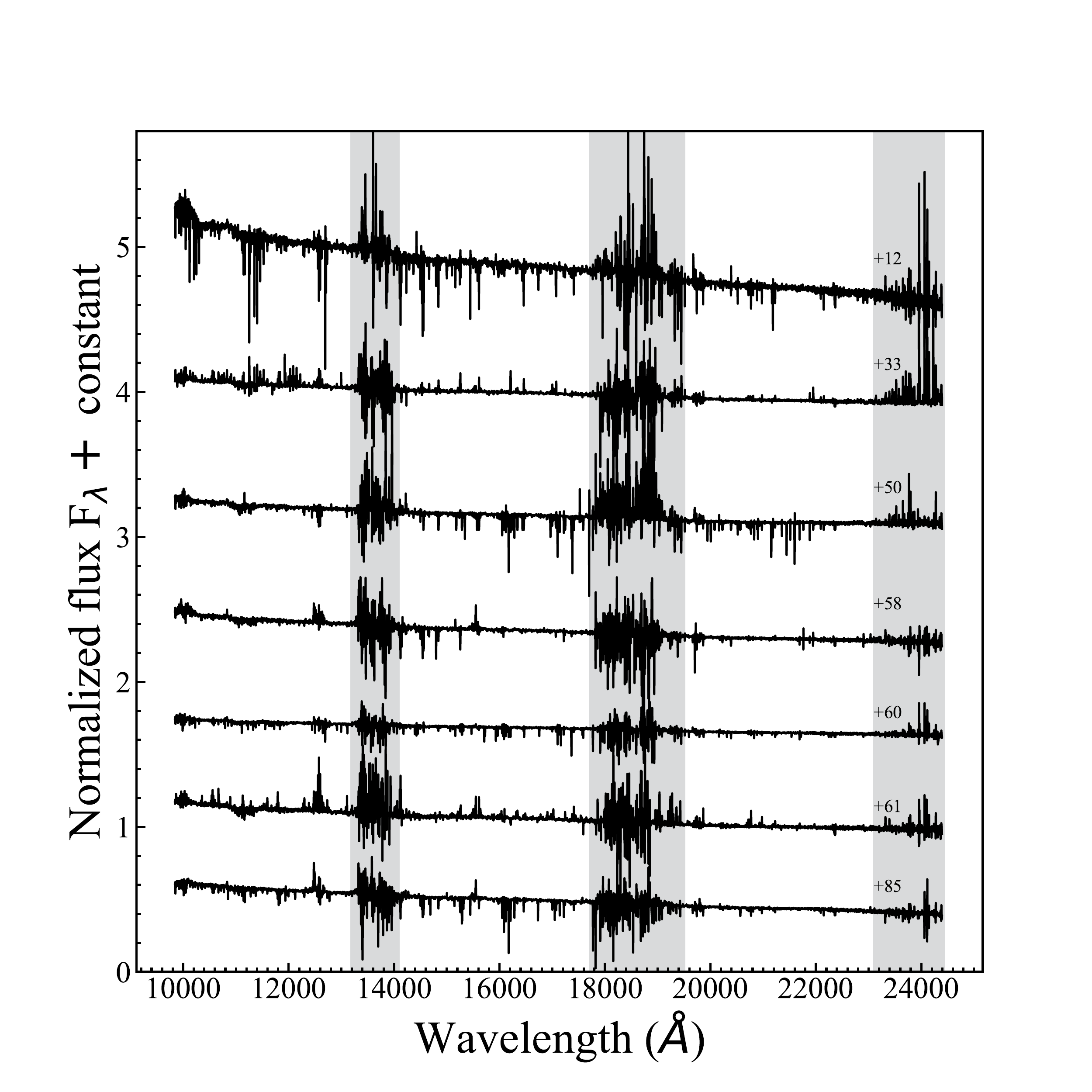}
\caption{Sequence of the rest-frame NIR spectra of iPTF16fnl taken with VLT/X-shooter. All the spectra have been corrected for reddening. The time since light curve peak of the observations is indicated. The area affected by telluric absorption are indicated by gray bands. No significant spectroscopic features have been detected.
}
\label{fig:NIRspec}
\end{figure*}

\begin{table*}
\centering
 \begin{minipage}{140mm}
 \caption{Properties of the narrow emission lines in X-shooter spectra } 
 \scriptsize
 \label{tbl:nlineXS}
 \begin{center}
 \begin{tabular}{@{}llcccccc}
\hline
\multicolumn{8}{c}{UVB arm}\\
\hline
MJD   &Phase [d] & \multicolumn{3}{c}{\ion{He}{II}} & \multicolumn{3}{c}{H$\upbeta$}\\
      &      & $\lambda$ [\AA]   & FWHM [\kms]     &EW [\AA] &$\lambda$ [\AA] & FWHM [\kms]  &EW [\AA]   \\
(1)   &(2)   & (3)               & (4)             & (5)     & (6)            & (7)          & (8)       \\
57\,644 &12    & --                & --              &--     &--          & --      &--    \\ 
57\,665 &33    & --                & --              &--     &--          & --      &--    \\  
57\,682 &50    &4685.65 $\pm$ 0.25 & 448 $\pm$ \phantom{1}41 &0.66 $\pm$ 0.08 &--          & --      &-- \\ 
57\,690 &58    &4683.46 $\pm$ 0.45 & 949 $\pm$ 101 &0.96 $\pm$ 0.14 &4864.14 $\pm$ 0.29 & 222 $\pm$ 42 & 0.17 $\pm$ 0.04\\
57\,692 &60    &4684.87 $\pm$ 0.39 & 836 $\pm$ \phantom{1}64   &0.72 $\pm$ 0.07&--          & --      &--   \\ 
57\,693 &61    &4685.06 $\pm$ 0.28 & 766 $\pm$ \phantom{1}54 & 0.69 $\pm$ 0.06 &--          & --      &-- \\ 
57\,717 &85    &4686.63 $\pm$ 0.28 & 340 $\pm$ \phantom{1}43 & 0.46 $\pm$ 0.07& 4863.44 $\pm$ 0.61 & 353 $\pm$ 89& 0.20 $\pm$ 0.07  \\ 
\hline
MJD   & Phase [d] &\multicolumn{3}{c}{\ion{O}{III} $\lambda$4959}& \multicolumn{3}{c}{\ion{O}{III} $\lambda$5007}\\
      &       &$\lambda$ [\AA] & FWHM [\kms] &EW [\AA]              &$\lambda$ [\AA] & FWHM [\kms]    &EW [\AA] \\
57\,644 & 12     &4957.81 $\pm$ 0.30 & 397 $\pm$ 44 & 0.31 $\pm$ 0.04 &5007.13 $\pm$ 0.07 & 293 $\pm$ 11 & 0.81 $\pm$ 0.04\\
57\,665 &33     &4959.76 $\pm$ 0.09 &  142 $\pm$ 12 & 0.27 $\pm$ 0.03& 5007.16 $\pm$ 0.07 & 268 $\pm$ 10 & 0.81 $\pm$ 0.04 \\
57\,682 &50     &4959.68 $\pm$ 0.20 &  165 $\pm$ 27 & 0.22 $\pm$ 0.05& 5007.43 $\pm$ 0.07 & 183 $\pm$ 10 & 0.70 $\pm$ 0.05 \\
57\,690 &58     & 4959.51 $\pm$ 0.20 & 191 $\pm$ 27 & 0.22 $\pm$ 0.04&5007.52 $\pm$ 0.09 &  292 $\pm$ 12 & 0.93 $\pm$ 0.05 \\
57\,692 &60     & 4959.62 $\pm$ 0.28 & 395 $\pm$ 40 & 0.37 $\pm$ 0.05& 5007.72 $\pm$ 0.06 & 236 $\pm$ \phantom{1}9  & 0.78 $\pm$ 0.04\\
57\,693 &61     &--                  & --           &--              & 5007.88 $\pm$ 0.06 & 331 $\pm$ \phantom{1}9  & 0.92 $\pm$ 0.03 \\
57\,717 &85     &4959.75 $\pm$ 0.34  & 269 $\pm$ 48 & 0.26 $\pm$ 0.06& 5007.95 $\pm$ 0.15 & 369 $\pm$ 21 & 0.93 $\pm$ 0.07 \\
\hline
\multicolumn{8}{c}{VIS arm}\\
\hline
MJD   &Phase [d] & \multicolumn{3}{c}{[\ion{N}{II}] $\lambda$6548} & \multicolumn{3}{c}{H$\upalpha$}\\
     &       &$\lambda$ [\AA]   & FWHM [\kms]    &EW [\AA]   &$\lambda$ [\AA]   & FWHM [\kms]    &EW [\AA] \\
57\,644 &12 & -- & -- & -- & 6554.10 $\pm$ 0.32 & 1010 $\pm$ 37 & 1.70 $\pm$ 0.08 \\
57\,665 &33 &-- & -- & --  &6560.95 $\pm$ 0.13 & \phantom{1}294 $\pm$ 16  & 1.29 $\pm$ 0.09 \\
57\,682 &50 &6547.81 $\pm$ 0.40 & 184 $\pm$ 49  & 0.15 $\pm$ 0.05& 6563.17 $\pm$ 0.11 & \phantom{1}344 $\pm$ 14 & 1.48 $\pm$ 0.08 \\
57\,690 &58 &-- & -- & --  &6563.44 $\pm$ 0.10 & \phantom{1}358 $\pm$ 12 & 1.74 $\pm$ 0.08 \\
57\,692 &60 &-- & -- & --  &6563.33 $\pm$ 0.05 & \phantom{1}322  $\pm$ \phantom{1}7 &1.27 $\pm$ 0.04 \\
57\,693 &61 &-- & -- & --  &6563.36 $\pm$ 0.07 & \phantom{1}328 $\pm$ \phantom{1}9 & 1.55 $\pm$ 0.05 \\
57\,717 &85 &6547.85 $\pm$ 0.26 & 595 $\pm$ 29 & 1.36 $\pm$ 0.08& 6563.67 $\pm$ 0.09 & \phantom{1}490 $\pm$ \phantom{1}9 & 3.07 $\pm$ 0.07 \\
\hline
MJD   &Phase [d] & \multicolumn{3}{c}{[\ion{N}{II}] $\lambda$6583} & \multicolumn{3}{c}{[\ion{S}{II}] $\lambda$6716}\\
     &       &$\lambda$ [\AA] & FWHM [\kms]  &EW [\AA]        &$\lambda$ [\AA]   & FWHM [\kms]    &EW [\AA] \\
57\,644 &12 & 6583.61 $\pm$ 0.07 & 119 $\pm$\phantom{1}7  & 0.36 $\pm$ 0.03&-- & -- & --\\
57\,665 &33 & 6583.01 $\pm$ 0.14 & 131 $\pm$ 15 & 0.39 $\pm$ 0.06&-- & -- & --\\
57\,682 &50 & 6584.76 $\pm$ 0.15 & 171 $\pm$ 17 & 0.35 $\pm$ 0.04&6709.17 $\pm$ 0.21 & 159 $\pm$ 21& 0.24 $\pm$ 0.04\\
57\,690 &58 & 6584.76 $\pm$ 0.12 & 294 $\pm$ 14 & 1.00 $\pm$ 0.06& 6708.87 $\pm$ 0.14 & 203 $\pm$ 14& 0.51 $\pm$ 0.05 \\
57\,692 &60 & 6584.99 $\pm$ 0.07 & 150 $\pm$ \phantom{1}8 & 0.30 $\pm$ 0.02& 6709.12 $\pm$ 0.11 & 207 $\pm$ 11 & 0.34 $\pm$ 0.02 \\
57\,693 &61 & 6585.24 $\pm$ 0.09 & 214 $\pm$ 10 & 0.58 $\pm$ 0.04&-- & -- & -- \\
57\,717 &85 & 6584.43 $\pm$ 0.10 & 318 $\pm$ 11 & 0.98 $\pm$ 0.04& 6709.30 $\pm$ 0.12 & 200 $\pm$ 13 & 0.42 $\pm$ 0.04 \\
\hline
\end{tabular}
\end{center}
Notes: Narrow emission lines properties inferred from the fit on the host-subtracted X-Shooter spectra. (1) MJD date of observations; (2)  Phase (days) with respect to the estimated date of TDE peak MJD 57\,632.1, according to \citealt{blagorodnova17}; (3) and (6) Central wavelength of the narrow component; (4) and (7) Full Width at Half Maximum of the narrow component; (5) and (8) Absolute value of the Equivalent Width of the narrow component.
\noindent
\end{minipage}
\end{table*} 

\bsp	
\label{lastpage}
\end{document}